\documentclass[a4paper,10pt]{article}
\title{State-space Correlations and Stabilities}
\author{}
\date{ Stefano Bellucci  \thanks{\noindent bellucci@lnf.infn.it} \ and
 Bhupendra Nath Tiwari \thanks{\noindent bntiwari.iitk@gmail.com}\\
\vspace{0.50 cm}
INFN-Laboratori Nazionali di Frascati\\
Via E. Fermi 40, 00044 Frascati, Italy.\\
\vspace{0.50 cm}
Department of Physics \\
Indian Institute of Technology-Kanpur,\\
Kanpur-208016, India.}
\vspace{0.50 cm}
\begin{document}
\maketitle
\abstract{ The state-space pair correlation functions and notion
of stability of extremal and non-extremal black holes in string
theory and M-theory are considered from the viewpoints of
thermodynamic Ruppeiner geometry. From the perspective of
intrinsic Riemannian geometry, the stability properties of these
black branes are divulged from the positivity of principle minors
of the space-state metric tensor. We have explicitly analyzed the
state-space configurations for (i) the two and three charge
extremal black holes, (ii) the four and six charge non-extremal
black branes, which both arise from the string theory solutions.
An extension is considered for the $D_6$-$D_4$-$D_2$-$D_0$
multi-centered black branes, fractional small black branes and two
charge rotating fuzzy rings in the setup of Mathur's fuzzball
configurations. The state-space pair correlations and nature of
stabilities have been investigated for three charged bubbling
black brane foams, and thereby the M-theory solutions are brought
into the present consideration. In the case of extremal black
brane configurations, we have pointed out that the ratio of
diagonal space-state correlations varies as inverse square of the
chosen parameters, while the off diagonal components vary as
inverse of the chosen parameters. We discuss the significance of
this observation for the non-extremal black brane configurations,
and find similar conclusion that the state-space correlations
extenuate as the chosen parameters are increased. In either of the
above configurations, notion of scaling property suggests that the
brane-brane pair correlations, which find an asymmetric nature in
comparison with the other state-space pair correlations, weaken
relatively faster and they relatively swiftly come into an
equilibrium statistical configuration. Novel aspects of the
state-space interactions may be envisaged from the coarse graining
counting entropy of underlying CFT microstates.}

\vspace{2cm}


strings, and related objects.\\

\newpage



\section{Introduction} 


State-space configurations involving thermodynamics of extremal and non-extremal black branes
in string theory \cite{9601029v2,9411187v3,9504147v2,0409148v2,9707203v1,0507014v1,0502157v4,
0505122v2} and $M$-theory \cite{0209114,0401129,0408106,0408122} possess rich intrinsic geometric
structures \cite{0606084v1,SST,BNTBull}. The present article thus focus attention on thermodynamic
perspectives of black branes, and thereby explicates nature of concerned state-space pair correlations
and associated stability of the solutions containing large number of branes and antibranes. Besides
several general notions which have earlier been analyzed in the condensed matter physics \cite{RuppeinerRMP,
RuppeinerA20,RuppeinerPRL,RuppeinerA27,RuppeinerA41}, we shall here consider specific string theory and
M-theory configurations thus mentioned with few thermodynamic parameters and analyze possible state-space
pair correlation functions and their scaling relations. Basically, the investigation which we shall follow
here entails certain intriguing features of underlying statistical fluctuations which can be defined in
terms of thermodynamic parameters. Given definite covariant state-space description of a consistent macroscopic
black brane solution, we shall expose (i) for what conditions the considered configuration is stable?,
(ii) how its state-space correlation functions scale in terms of the chosen thermodynamic parameters?
In this process, we shall also enlist complete set of non-trivial relative state-space correlation
functions of the configurations considered in \cite{BNTBull}. It may further be envisaged in this
direction that the similar considerations indeed remain valid over the other black holes in general
relativity \cite{gr-qc/0601119v1, gr-qc/0512035v1, gr-qc/0304015v1, 0510139v3}, attractor black holes \cite{9508072v3,9602111v3,new1,new2, 0702019v1,0805.1310} and Legendre transformed finite parameter
chemical configurations \cite{Weinhold1, Weinhold2}, quantum field theory and QCD backgrounds \cite{BNTSBVC}.

On other hand, the state-space configurations of four dimensional $\mathcal N=2$ black holes can be
characterized by electric and magnetic charges $q_J$ and $ p^I $ which arise from usual flux integrals of
the field strength tensors and their Poincar\'{e} duals. In such cases, the near horizon geometry of an
extremal black hole turns out to be an $AdS_2 \times S^2 $ manifold which describes the Bertotti-Robinson
vacuum associated with the black hole. The area of the black hole horizon $A$ and thus the macroscopic
entropy \cite{9508072v3, 9602111v3, new1, new2} is given as $S_{macro}= \pi \vert Z_{\infty} \vert^2$.
Such attractor solutions and their critical properties have further been explored from an effective
potential defined in $\mathcal N=2,D=4$ supergravities coupled to $n_{V}$ abelian vector multiplets
for an asymptotically flat extremal black holes describing $(2n_{V}+2)$-dyonic charges and $n_{V}$
number of complex scalar fields which parameterize $n_{V}$-dimensional special K\"{a}hler manifold
\cite{Bull1,Bull2,Bull3,0702019v1}. The statistical entropy of the supersymmetric charge black holes
coming from counting the degeneracy of bound states have been examined against the macroscopic Wald
entropy \cite{gr-qc/9307038, gr-qc/9502009, 9305016} which further agrees term by term with the higher
derivative supergravity corrections, as well \cite{0505122v2}. In order to study the respective cases
of non-extremal black branes of such black holes, one may add some total mass or corresponding antibranes
to the chosen extremal black brane configuration, and thereby possible specific computation of the black brane
entropy can either be performed in the macroscopic setup, or that of associated microscopic considerations.
Furthermore, the investigation of \cite{9602043v2,0412287} shows a match between the $ S_{micro}$ and
$ S_{macro}$ for the non-extremal configurations carrying definite brane and antibrane charges.

There has been various extremal black holes \cite{9309145v2,9405117v1,9411187v3,9504147v2,0505122v2}
and non-extremal black brane space-times \cite{9602043v2}, multi-centered black brane configurations
\cite{0705.2564v1, 0702146v2}, small black holes with fractional branes \cite{9707203v1,0507014v1,0502157v4,
0505122v2}, fuzzy rings in Mathur's fuzzballs as well as the subensemble theoretic set-up \cite{0109154v1,
0202072v2, 0706.3884v1, 0804.0552} for which the notion of state-space geometry has been introduced in
\cite{BNTBull}. Similar properties have further been explored for the three charge bubbling black brane
solutions in M-theory as well \cite{0604110}. We shall thus systematically present the status of lower
dimensional black hole thermodynamic configurations from the view-points of an intrinsic Riemannian geometry.
In this connection, the microscopic perspective of black branes have been analyzed in \cite{0412322,0409148v2}
and their thermodynamic geometry whose basics have been motivated in \cite{RuppeinerPRL,RuppeinerPRD78} has
recently been investigated \cite{BNTBull}. We shall show that the similar phenomena may further be explored
to exhibit how state-space local correlations scales and under what conditions a chosen black brane solution
correspond to stable state-space configuration. In this case, it has been possible to out-line that there is an
explicit correspondence between the parameters of microscopic spectrum and macroscopic properties of a class of
extremal \cite{9309145v2,9405117v1} and associated non-extremal black brane systems \cite{9711200v3}. It has there
after been pointed out that there should exist definite microscopic origins of underlying statistical and thermodynamic
interactions among the microstates of black brane configurations which give rise to an intrinsic Riemannian geometry.
In turn, we apperceive from \cite{9601029v2} that such notions may in turn be investigated from the perspective of
statistical fluctuations which arise from the coarse graining entropy of chosen configuration. 

The state-space geometry thus defined introduces in particular that the thermodynamic interactions
considered as the function of charges, angular momenta, and mass of a given black brane configuration
may be characterized by an ensemble of equilibrium microstates of underlying microscopic configurations.
Furthermore, it has been observed in all such cases that there exist a clear mechanism on the black brane
side that describes the notion of interactions on the state-space which turn out to be regular intrinsic
Riemannian manifold or vice-verse. In fact, one may therefore exhibit well-defined intrinsic thermodynamic
geometries associated each other via conformal transformation. The physical observations thus found are
consistent with existing picture of the microscopic CFT \cite{9711200v3,0206126,0212204,Bekenstein} that the
microscoipc entropy $S_{micro}$ counts the states of black brane configuration in a field theory description
dual to the gravitational description. In fact, such conventional understanding of the entropy is bases on coarse
graining over a large number of microstates \cite{0202072v2,0502050v1} and thus it turns out to be a crucial
ingredient in realizing equilibrium macroscopic thermodynamic geometry. It has been shown \cite{RuppeinerRMP,
RuppeinerPRD78,BNTBull} that the components of state-space metric tensor defined as Hessian matrix of the entropy
signify state-space pair correlation functions and the associated state-space curvature scalar implies the nature
of global correlation volume of the underlying statistical system. Such intrinsic geometric local and global correlations
have initially been studied for the thermodynamic configurations of general relativity black holes \cite{gr-qc/0601119v1, gr-qc/0512035v1,gr-qc/0304015v1,0510139v3}, and there after they have been brought into focus to the string theory
or M-theory black holes \cite{SST, 0801.4087v1}. Furthermore, it turns our that the state-space interactions
generically remain finite and non-zero when small thermal fluctuations in the canonical ensemble are taken into
account \cite{0606084v1}.

As mentioned above, we shall analyze the intrinsic geometric relative correlations and notion of stability for
large class of extremal as well as non-extremal black hole configurations at an attractor fixed point solution.
The study of chosen thermodynamic systems has rather been already intimated in \cite{BNTBull}, and in this paper
our specific goal shall thereby be to explicate their further state-space properties for large charge rotating and non-rotating
black brane configurations. A similar application has indeed been performed for the spherical horizon and non-spherical horizon M-theory configurations, see for details \cite{RotBHs,BSBRs}. Moreover, it has been demonstrated that various state-space geometric notions turn out to be well-defined, even at zero temperature \cite{0801.4087v1}. Interestingly, such geometric studies in connectiom with attractor fixed point(s) are expected to give further motivations for analyzing large class of extremal and non-extremal black brane configurations under Plank length corrections \cite{BNTuncertainity} or that of the higher derivative stringy $\alpha^{\prime}$-corrections \cite{gr-qc/9307038, gr-qc/9502009,9305016, 0801.4087v1} being incorporated via definite Sen entropy functions \cite{Sen1,Sen2,Sen3,Sen4,Sen5,Sen6,Sen7,Sen8} which are obtained over an underlying supergravity effective action for a given moduli space configuration. These notions shall however be left for future investigation.

The present article is organized as follows. The first section motivates why to study state-space
configurations of the string theory and M-theory black brane solutions. In either of the subsequent
state-space configurations, we shall analyze scaling properties of the state-space pair correlation
functions, possiblepositivity of heat capacities and non-trivial state-space stabilities. In section 2,
we introduce state-space properties for two charge extremal black holes or an excite strings carrying
$ n_1 $ number of winding modes and $ n_p $ number of momentum modes. In section 3, we demonstrate
that the nature of state-space correlations find similar pattern for the three charge extremal black holes.
In section 4, we consider state-space analysis for the non-extremal black branes corresponding to three
charge $D_1$-$D_5$-$P$ extremal solutions with an addition of antibrane charge. In section 5, we show
explicitly that the similar conclusions hold for the six charge non-extremal black branes, as wewll.
In section 6, we focus on the state-space correlations of multi-centered $D_6$-$D_4$-$D_2$-$D_0$ configurations,
and thereby expose the respective cases for the single center and double center four charge solutions. In section
7, we demonstrate state-space correlation properties for the two cluster, three cluster and then arbitrary
finite cluster $D_0$-brane fractionations in the $D_0$-$D_4$ black branes. In section 8, we explicate that
the similar state-space geometric notions hold for three parameter fuzzy rings introduced in the set-up of
Mathur's fuzzball solutions. In section 9, we extend our state-space analysis to the three charge bubbling
black brane solutions in M-theory. Finally, section 10 contains some concluding issues and the other
implications of the state-space geometry of string theory and M-theory black brane solutions for future.

\section{Two charge extremal configurations}

In this section, we analyze the non-trivial state-space interactions among various
microstates of given black brane configurations, and thereby explore present consideration
from the perspective of relative correlations and definite state-space stabilities.
To illustrate the basic idea of state-space geometry of string theory black holes,
we first consider the case of simplest example describing two charge extremal
configurations. It turns out that the state-space geometry of such configurations
may be analyzed in terms of the winding modes and the momentum modes of an excited
string carrying $ n_1 $ number of winding modes and $ n_p $ number of momentum modes.
To be concrete, we consider the study of state-space geometry arising from an extremal
black hole whose microstates are characterized by the momentum and winding numbers, and
microscopic entropy formula \cite{9309145v2,9405117v1,9411187v3,9504147v2} obtained from
large charge degeneracy of states reduces to
\begin{equation}
S_{micro}= 2\sqrt{2n_1 n_p}
\end{equation}

Macroscopically, the entropy of such two charged black holes may be computed
by considering the electric magnetic charges on the $D_4$ and $D_0$ branes,
with ascertained compactifications to obtain $ M_{3,1} $ black hole space-time.
There certainly exist higher derivative corrections in string theory, like
for instance the $R^2$ corrections or $ R^4$-corrections to the standard
Einstein action, and thus these corrections make the horizon area non-zero,
as the horizon of vanishing Bekenstein-Hawking entropy black holes is being
stretched by such higher derivative $\alpha^{\prime}$ corrections. The
computation of corresponding macroscopic entropy is usually accomplished by
assuming spherically symmetric ansatz for the non-compact spatial directions
\cite{0409148v2}. On other hand, the microscopic entropy may whereas be counted
by considering an ensemble of weakly interacting D-branes \cite{9512078v1}.
One indeed finds for $n_4$ number of $D_4$ branes and $n_0$ number of $D_0$
branes that the both entropy do match with
\begin{equation}
S_{micro}= 2 \pi \sqrt{n_0 n_4}= S_{macro}
\end{equation}

First of all, an immediate goal would be to understand state-space geometric
notions associated with the leading order two charge black brane solutions,
which we shall thus consider via an analysis of the state-space configurations
of either an excited string or that of the $D_0$, $D_4 $ black holes. The analysis
follows directly by computing the Hessian matrix of the entropy with respect
to concerned extensive thermodynamic variables of the either configurations.
It is worth to mention that the respective  entropy can simply be defined as
a function of the winding and momentum charges of the string or that of the
two charge $D_0, D_4 $ black branes. As such a configuration is uniquely
related to each other and have the same expressions for their entropy. Thus,
we focus our attention on the two charge $D_0$-$D_4$ configurations.

From the given expression of the entropy of two charge $D_0$-$D_4$ configuration,
we observe that the statistical pair correlations may easily be accounted by
simple geometric descriptions being expressed in terms of the brane numbers connoting
an ensemble of microstates of the $D_0$-$D_4$ black hole solutions. Furthermore,
it is not difficult to see that the components of the state-space metric tensor
describing equilibrium statistical pair correlations may be computed from the
negative Hessian matrix of the entropy. As an easy result, we deduce for all
allowed values of the parameters of the two charge $D_0$-$D_4$ black holes that
the components of underlying state-space metric tensor\footnote{ In the present
and subsequent sections, we shall invariably use for given set of brane and antibrane
charges and angular momenta $X_a= (X_1,X_2,\cdots, X_k) \in M_k $ that the tensor
notations $g_{X_iX_j}$ and $g_{ij}$ signify the same intrinsic state-space object.}
are given as
\begin{eqnarray}
g_{n_0n_0}= \frac{\pi}{2 n_0} \sqrt{\frac{n_4}{n_0}}, \ 
g_{n_0n_4}= -\frac{\pi}{2} \frac{1}{\sqrt{n_0n_4}}, \ 
g_{n_4n_4}= \frac{\pi}{2 n_4} \sqrt{\frac{n_0}{n_4}}
\end{eqnarray}

It is thus evident that the principle components of state-space metric tensor
$\lbrace g_{n_in_i} \vert \ i=0,4 \rbrace$ essentially signify a set of definite
heat capacities (or the related compressibilities) whose positivity in turn apprises
that the $D_0$-$D_4$ black brane solutions comply an underlying equilibrium statistical
configuration. In particular, it is also clear for an arbitrary number of $D_0$ and
$D_4$ branes that the associated state-space metric constraints as the diagonal
pair correlation functions remain positive definite, \textit{viz.}, we have
\begin{eqnarray}
g_{n_in_i}&>& 0 \ \forall \ i \in \{0,4 \} \mid n_i > 0 
\end{eqnarray}

The case of finitely many $D_0$-$D_4$ branes indeed agrees with an
expectation that the non diagonal component $g_{n_0n_4}$ of the
state-space metric tensor respectively finds some non-zero
negative value. Furthermore, we visualize from the definition of
state-space metric tensor that the ratios of principle components
of Gaussian statistical pair correlations vary as inverse square
of the concerned brane charges; while that of the off-diagonal
correlations modulate only inversely. Interestingly from just
designated state-space pair correlations of these two charge black
hole configurations, it follows for distinct $i,j \in \lbrace 0,4
\rbrace $ that the following expressions define possible set of
admissible scaling relations
\begin{eqnarray}
\frac{g_{ii}}{g_{jj}}= (\frac{n_j}{n_i})^2, \ 
\frac{g_{ij}}{g_{ii}}= -\frac{n_i}{n_j}
\end{eqnarray}

In order to determine the global properties of fluctuating two charge $D_0$-$D_4$
extremal configurations, we need to determine stabilities along each intrinsic directions,
each intrinsic planes, and intrinsic hyper-planes, if any, as well as on the full intrinsic
state-space manifold. Nevertheless, we notice that the underlying state-space manifold
in the present case is just an ordinary intrinsic surface, and thus the set of stability
criteria on various possible state-space configurations could simply be determined by the two
possible principle minors, \textit{viz.}, $p_1$ and $p_2$. For all $n_0$ and $n_4$, we find
that the first minor constraint $p_1>0$ directly follows from the positivity of the first
component of metric tensor
\begin{eqnarray}
p_1&=& \frac{\pi}{2 n_0} \sqrt{ \frac{n_4}{n_0}}
\end{eqnarray}

Whereas, the minor constraint, $p_2>0$ becomes the positivity of the determinant of metric
tensor which nevertheless vanishes identically for all allowed values of the $n_0$ and $n_4$.
In this case, we explicitly see that the minor constraints is not fulfilled, \textit{viz.},
the minor $p_2:= g(n_0,n_4)$ takes the null value, and thus the leading order consideration
of degeneracy of the states of large charge $D_0$-$D_4$ extremal black branes or excited
strings with $n_1$ number of winding and $n_p$ number of momenta find degenerate intrinsic
state-space configurations.

\section{Three charge extremal configurations}

To have test of a more charged black hole state-space
configuration, we may add $ n_5 $ number of $ D_5 $ branes to the
above excited string configuration, and then one finds that the
leading order entropy of three charge extremal black hole may be
obtained from the two derivative level Einstein-Hilbert action. It
is well-known \cite{9601029v2} that the entropy of extremal
$D_1$-$D_5$ solutions arising from Einstein-Hilbert action is
proportional to the area of the horizon and the corresponding
microscopic entropy may as well be counted by considering an
ensemble of weakly interacting D-branes. It turns out that the two
entropies match and they take the following form:
\begin{eqnarray}
S_{micro}= 2 \pi \sqrt{n_1 n_5 n_p}= S_{macro}
\end{eqnarray}

The state-space geometry describing the correlations between the equilibrium microstates
of the three charged rotating extremal $D_1$-$D_5$ black holes resulting from the degeneracy
of the microstates may easily be computed as earlier from the Hessian matrix of the entropy
with respect to the number of $D_1$, $D_5$ branes and Kaluza-Klein momentum, {\it viz,}, $n_1$,
$n_5$ and $n_p$. We thence see that the components of state-space metric tensor are given by
\begin{eqnarray}
g_{n_1n_1}&=& \frac{\pi}{2n_1} \sqrt{\frac{n_5 n_p}{n_1}}, \ 
g_{n_1n_5}= -\frac{\pi}{2} \sqrt{\frac{n_p}{n_1 n_5}} \nonumber \\
g_{n_1n_p}&=& -\frac{\pi}{2} \sqrt{\frac{n_5}{n_1 n_p}}, \ 
g_{n_5n_5}= \frac{\pi}{2n_5} \sqrt{\frac{n_1 n_p}{n_5}} \nonumber \\
g_{n_5n_p}&=& -\frac{\pi}{2} \sqrt{\frac{n_1}{n_5 n_p}}, \ 
g_{n_pn_p}= \frac{\pi}{2n_p} \sqrt{\frac{n_1 n_5}{n_p}}
\end{eqnarray}

The statistical pair correlations thus ascertained could in turn be accounted by
simple microscopic descriptions being expressed in terms of the number of $D_1$-$D_5$
branes and Kaluza-Klein momentum connoting an ensemble of microstates of the extremal
black hole configurations. Furthermore, it is evident that the principle components
of the pair correlation functions remain positive definite for all the allowed values
of concerned three parameters of the black holes. As a result, we thus easily observe
that the concerned state-space metric constraints are satisfied with
\begin{eqnarray}
g_{n_in_i}&>& 0 \ \forall \ i \in \{1,5,p\} \mid n_i > 0
\end{eqnarray}

We thus see in this case that the principle components of state-space metric tensor
$\lbrace g_{n_i n_i}, g_{n_pn_p} \vert \ i=1,5 \rbrace$ essentially signify a set of
definite heat capacities (or related compressibilities) whose positivity demonstrates
that the three charge $D_1$-$D_5$-$P$ black holes comply underlying locally stable
equilibrium statistical configuration. Furtheremore, we inspect that an addition of
Kaluza-Klein momentum charge do not alter the conclusion of excited string system
that the $D_1$-$D_5$-$P$ configuration with finitely many $D_1$-$D_5$ branes and
momentum excitations agrees with our naive expectation that respective non diagonal
components, \textit{viz.}, $g_{ij}$ and $g_{ip}$ of the state-space metric tensor
can find some non-positive values.

Interestingly, the ratios of the principle components of metric tensor describing Gaussian
statistical pair correlations vary as inverse square of the brane numbers and momentum charge;
while that of the off-diagonal rations of the state-space correlations modulate only inversely.
It further follows from the above expressions that we may explicitly visualize for distinct
$i,j \in \lbrace 1,5 \rbrace $ and $p$ that the list of relative correlation functions thus
described is consisting of the following scaling properties
\begin{eqnarray}
\frac{g_{ii}}{g_{jj}}&=& (\frac{n_j}{n_i})^2, \ 
\frac{g_{ii}}{g_{pp}}= (\frac{n_p}{n_i})^2, \ 
\frac{g_{ii}}{g_{ij}}= -(\frac{n_j}{n_i}) \nonumber \\
\frac{g_{ii}}{g_{ip}}&=& -(\frac{n_p}{n_i}), \ 
\frac{g_{ip}}{g_{jp}}= (\frac{n_j}{n_i}), \ 
\frac{g_{ii}}{g_{jp}}= -(\frac{n_jn_p}{n_i^2})\nonumber \\
\frac{g_{ip}}{g_{pp}}&=& -(\frac{n_p}{n_i}), \ 
\frac{g_{ij}}{g_{ip}}= (\frac{n_p}{n_j}), \ 
\frac{g_{ij}}{g_{pp}}= -(\frac{n_p^2}{n_in_j})
\end{eqnarray}

Along with the positivity of principle components of state-space metric tensor,
we need to demand in order to accomplish the local stability of associated system
that all the principle minors should be positive definite. It is nevertheless not
difficult to compute the principle minors of the Hessian matrix of the entropy of
three charge $D_1$-$D_5$-$P$ extremal black holes. In fact, after some manipulations
one encounters that the local stability conditions along the principle line and that
of respective two dimensional surface of concerned state-space manifold be simply
measured by the following equations
\begin{eqnarray}
p_1= \frac{\pi}{2n_1} \sqrt{\frac{n_5 n_p}{n_1}}, \ 
p_2= -\frac{\pi^2}{4n_1n_5^2n_p} (n_p^2 n_1+ n_5^3)
\end{eqnarray}

For all physically allowed values of brane numbers and momentum charge of the $D_1$-$D_5$-$P$
extremal black holes, we thus notice that the minor constraint $p_2(n_1, n_5, n_p)>0$ never
gets satisfied for any real positive physical parameters. In particular, we may easily inspect
that the nature of state-space geometry for the three charge $D_1$-$D_5$-$P$ extremal black holes
is that these solutions are stable along the line on state-space, but have planer instabilities.
It is easy to stipulate that our conclusion holds for arbitrary number of $D_1$-$D_5$ branes
and Kaluza-Klein momentum.

In the view-point of the simplest two charge extremal solutions, it tuns out that the local
stability on the entire equilibrium phase-space configurations of the $D_1$-$D_5$-$P$ extremal
black holes may clearly be determined by computing the determinant of the underlying state-space
metric tensor. As in the previous example, it is easy to observe that the state-space metric tensor
is a non-degenerate and everywhere regular function of the brane charges, $n_1$ and $n_5$ and
Kaluza-Klein momentum charge $n_p$. In particular, we find the under present consideration that
the determinant of the metric tensor as the highest principle minor $p_3:=g$ of the Hessian matrix
of the entropy obtains a simple form \begin{eqnarray}
\Vert g \Vert= -\frac{1}{2}\pi^3 (n_1 n_5 n_p)^{-1/2}
\end{eqnarray}

Moreover, we observe that the determinant of the metric tensor does not take a positive
definite, well-defined form, and thus there is no positive definite globally well-defined
volume form on the state-space manifold $(M_3,g)$ of concerned three charge $D_1$-$D_5$-$P$
extremal system. In turn, the non-zero value of the determinant of state-space metric tensor
$ g(n_1,n_5,n_p) $ indicates that the extremal $D_1$-$D_5$-$P$ solution may decay into some
other degenerate vacuum state configurations procuring the same corresponding entropy or
microscopic degeneracy of states. Here, we further notice independent of the microscopic
type-II string description or heterotic string description that the three charge $D_1$-$D_5$-$P$
black holes when considered as a bound state of the $D_1$-$D_5$-brane microstates and Kaluza-Klein
excitations do not correspond to an intrinsically stable statistical configuration. It is worth
to mention as introduced in \cite{BNTBull,0801.4087v1} in order to divulge phase transitions
and related global state-space properties that this statistical system remains everywhere regular
as long as the number of brane and Kaluza-Klein momentum charge take finite values.

In the next two sections, we shall deal with the state-space geometry of non-extremal black branes
in string theory with two/ three charges and two/ three anticharges of leading order entropy configurations.
After defining state-space metric tensor, we shall analyze scaling properties of possible state-space
pair correlation functions and stability requirements for chosen non-extremal black brane solution.

\section{Four charge non-extremal configurations}

The present section examines state-space configuration of the non-extremal black holes,
and extends our intrinsic geometric assessments for the $D_1$-$D_5$ black holes having
non-zero momenta along the clockwise and anticlockwise directions of Kaluza-Klein compactification
circle $ S^1$. For the purpose of critical ratifications, we shall focus our attention on the
state-space geometry arising from the entropy of non-extremal black hole, which one can simply
achieve just by adding corresponding anti-branes to the chosen extremal black brane solution.
What follows in precise that we shall first consider the simplest example of such systems, \textit{viz.},
a string having large amount of winding and $D_5$ brane charges: $n_1, n_5$ with some extra energy,
which in the microscopic description creates an equal amount of momenta running in opposite directions
of the $ S^1$. In this case, the entropy has been calculated from both the microscopic and macroscopic
perspective \cite{9602043v2}, and matches for given total mass and brane charges. In particular,
it has been shown in \cite{9602043v2} that the either of above entropies satisfy
\begin{eqnarray}
S_{micro}= 2 \pi \sqrt{n_1 n_5} (\sqrt{n_p}+ \sqrt{ \overline{n_p}})= S_{macro}
\end{eqnarray}

We thence analyze that the state-space covariant metric tensor defined as negative Hessian
matrix of entropy with respect to number of $D_1$, $D_5$ branes $\{ n_i \mid i= 1, 5 \} $ and
opposite Kaluza-Klein momentum charges $\{n_p, \overline{n_p} \}$\footnote{ In this section,
the notations $\overline{n_p}$ and $n_{\overline{p}}$ shall imply the same Kaluza-Klein momentum
charges which are in opposite direction of the $n_p$ momentum charge, and flow along the $ S^1$.}.
The associate components of the state-space metric tensor and stability parameters are thus easy
to compute for the non-extremal $D_1$-$D_5$ black holes. In fact, a direct computation finds
that the components of the metric tensor take the following expression
\begin{eqnarray}
g_{n_1 n_1}&=&  \frac{\pi}{2}\sqrt{\frac{n_5}{n_1^{3}}}(\sqrt{n_p}+{\sqrt{\overline{n_p}}}), \ 
g_{n_1 n_5}= -\frac{\pi}{2\sqrt{n_1 n_5}} (\sqrt{n_p}+{\sqrt{\overline{n_p}}})\nonumber \\
g_{n_1 n_p}&=& -\frac{\pi}{2} \sqrt{\frac{n_5}{n_1 n_p}}, \ 
g_{n_1 \overline{n_p}}= -\frac{\pi}{2} \sqrt{\frac{n_5}{n_1 \overline{n_p}}} \nonumber \\
g_{n_5 n_5}&=&  \frac{\pi}{2}\sqrt{\frac{n_1}{n_5^{3}}}(\sqrt{n_p}+{\sqrt{\overline{n_p}}}), \ 
g_{n_5 n_p}= -\frac{\pi}{2} \sqrt{\frac{n_1}{n_5 n_p}} \nonumber \\
g_{n_5 \overline{n_p}}&=& -\frac{\pi}{2} \sqrt{\frac{n_1}{n_5 \overline{n_p}}}, \ 
g_{n_p n_p}= \frac{\pi}{2} \sqrt{\frac{n_1 n_5}{n_p^{3}}} \nonumber \\
g_{n_p \overline{n_p}}&=& 0, \ 
g_{\overline{n_p} \overline{n_p}}= \frac{\pi}{2} \sqrt{\frac{n_1 n_5}{\overline{n_p}^{3}}}
\end{eqnarray}

It is clear that there exist an intriguing intrinsic geometric enumeration which describes
possible nature of statistical pair correlations. The present framework affirms in turn that
the concerned state-space pair fluctuations determined in terms of the brane and anti-brane
numbers (or brane charges) of the $D_1$-$D_5$-$P$ non-extremal black holes demonstrate definite
expected behavior of the underlying heat capacities. Hitherto, we see apparently that the
principle components of statistical pair correlations remain positive definite quantities for
all admissible values of underlying configuration parameters of the black branes. It may easily
be observed that the following state-space metric constraints are satisfied
\begin{eqnarray}
g_{n_in_i}> 0 \ \forall \ i= 1,5; \ 
g_{n_an_a}> 0 \ \forall \ a= p, \overline{p}
\end{eqnarray}

We thus physically note that the principle components of the state-space metric tensor
$\lbrace g_{n_in_i}, g_{n_an_a} \ \vert \ i=1,5; a= p, \overline{p} \rbrace$ signify a
set of heat capacities (or the associated compressibilities) whose positivity exhibits
that the underlying black hole system is in locally equilibrium statistical configuration
of the branes and anti-branes. The present analysis thus complies that the positivity of
$g_{n_an_a}$ obliges that the dual conformal field theory living on the boundary must prevail
an associated non-vanishing value of the momentum charges associated with large integers
$n_p, \overline{n_p}$ defining the degeneracy of microscopic conformal field theory.

It follows from the above expressions of the components of state-space metric tensor
that the ratios of principle components of statistical pair correlations vary as inverse
square of the brane numbers; while one finds in specific limit of leading order entropy
that the off-diagonal correlations vary only inversely. Interestingly, we may as a sequel
visualize for the distinct $i, j \in \{1,5\}$, and $k, l \in \{p,\overline{p}\}$ describing
four charge non-extremal $D_1$-$D_5$-$P$-$\overline{P}$ black holes that the statistical
pair correlations thus proclaimed consist the following set of scaling relations
\begin{eqnarray}
\frac{g_{ii}}{g_{jj}}&=& (\frac{n_j}{n_i})^2, \ 
\frac{g_{ii}}{g_{kk}}= \frac{n_k}{n_i^2} \sqrt{n_k}(\sqrt{n_p}+\sqrt{ \overline{n_p}}), \ 
\frac{g_{ii}}{g_{ij}}= -\frac{n_j}{n_i} \nonumber \\
\frac{g_{ii}}{g_{ik}}&=& -\frac{\sqrt{n_k}}{n_i}(\sqrt{n_p}+\sqrt{ \overline{n_p}}), \ 
\frac{g_{ik}}{g_{jk}}= \frac{n_j}{n_i}, \ 
\frac{g_{ii}}{g_{jk}}= -\frac{n_j}{n_i^2} \sqrt{n_k}(\sqrt{n_p}+\sqrt{ \overline{n_p}}) \nonumber \\
\frac{g_{ik}}{g_{kk}}&=& -\frac{n_k}{n_i}, \ 
\frac{g_{ij}}{g_{ik}}= \frac{\sqrt{n_k}}{n_j} (\sqrt{n_p}+\sqrt{ \overline{n_p}}), \ 
\frac{g_{ij}}{g_{kk}}= - \frac{n_k}{n_in_j} \sqrt{n_k}(\sqrt{n_p}+\sqrt{ \overline{n_p}})
\end{eqnarray}

We further see that the list of other relative correlation functions concerning the non-extremal
$D_1$-$D_5$-$P$-$\overline{P}$ black holes are
\begin{eqnarray}
\frac{g_{ik}}{g_{il}}&=& \sqrt{\frac{n_l}{n_k}}, \ 
\frac{g_{ik}}{g_{jl}}= \frac{n_j}{n_i} \sqrt{\frac{n_l}{n_k}}, \ 
\frac{g_{kl}}{g_{ij}}= 0 \nonumber \\
\frac{g_{kl}}{g_{ii}}&=& 0, \ 
\frac{g_{kk}}{g_{ll}}= (\frac{n_l}{n_k})^{3/2}, \ 
\frac{g_{kl}}{g_{kk}}= 0
\end{eqnarray}

To investigate the entire set of geometric properties of fluctuating non-extremal $D_1$-$D_5$
configurations, we need to determine stability along the each intrinsic directions, each intrinsic
planes, as well as on the full intrinsic state-space manifold. Here, we may adroitly compute the
principle minors from the Hessian matrix of associated entropy concerning the four charge string
theory non-extremal black hole solution carrying $D_1$, $D_5$ charges and left and right KK momenta.
In fact, a simple manipulations discovers that the set of local stability criteria on various
possible surfaces and hyper-surfaces of the underlying state-space configuration are respectively
determined by the following set of equations
\begin{eqnarray}
p_0&=& 1 , \ 
p_1= \frac{\pi}{2}\sqrt{\frac{n_5}{n_1^{3}}}(\sqrt{n_p}+{\sqrt{\overline{n_p}}}) \nonumber \\
p_2&=& 0, \ 
p_3= -\frac{1}{2n_p}\frac{\pi^3}{\sqrt{n_1n_5}}(\sqrt{n_p}+\sqrt{\overline{n_p}})
\end{eqnarray}

For all physically admitted values of the brane and antibrane charges (or concerned brane numbers)
of the non-extremal $D_1$-$D_5$ black holes, we can thus easily ascertain that the minor constraint,
{\it viz.}, $p_2(n_i, n_p,\overline{n_p})=0$ exhibits that the two dimensional state-space configurations
are not stable for any value of the brane numbers and assigned Kaluza-Klein momenta. Similarly, the
positivity of $p_1(n_i, n_p,\overline{n_p})$ for arbitrary number of branes shows that the underlying
fluctuating configurations are locally stable because of the line-wise positive definiteness.

The constraint $p_3(n_i, n_p,\overline{n_p})>0$ respectively imposes the condition that the system may
ever not attain stability on three dimensional subconfigurations for all given positive Kaluza-Klein
momenta and given positive $n_i$'s. In particular, these constraints enable us to investigate potential
nature of the state-space geometric stability for leading order non-extremal $D_1$-$D_5$ black branes.
We thus observe that the presence of planer and hyper-planer instabilities exist for the spherical
horizon non-extremal $D_1$-$D_5$ solutions. We expect altogether in the view points of subleading higher
derivative contributions in the entropy that the involved systems demand some restriction on allowed
value of the Kaluza-Klein momenta and number of branes and antibranes.

Moreover, it is not difficult to enquire the complete local stability of the full state-space
configuration of non-extremal $D_1$-$D_5$ black branes, and in fact it may simply be acclaimed by
computing the determinant of the state-space metric tensor. Nevertheless, it is possible to enumerate
compact formula for the determinant of the metric tensor. For the different allowed values of brane
numbers, {\it viz.}, $\{n_1,n_5 \}$ and Kaluza-Klein momenta $\{n_p,\overline{n_p} \}$, one apparently
discovers from concerned intrinsic geometric analysis that the non-extremal $D_1$-$D_5$ system admits
the following expression for the determinant of the state-space metric tensor
\begin{eqnarray}
g(n_1, n_5, n_p,\overline{n_p})= -\frac{1}{4}\frac{\pi^4}{(n_p\overline{n_p})^{3/2}}(\sqrt{n_p}+\sqrt{\overline{n_p}})^2
\end{eqnarray}

Furthermore, we may exhibit that the nature of the statistical interactions and the other global
properties of the $D_1$-$D_5$ non-extremal configurations are indeed not really perplexing to anatomize.
In this regard, one computes certain global invariants of the state-space manifold $(M_4,g)$ which
in the present case can easily be determined in terms of the parameters of underlying brane configurations.
Here, we may work in the large charge limit in which the asymptotic expansion of the entropy of non-extremal
$D_1$-$D_5$ system is valid. In particular, we notice that the state-space scalar curvatureas indicated in
\cite{BNTBull} generically remains non vanishes for all finite value of the brane charges and Kaluza-Klein
momenta. Thus for physically acceptable parameters, the large charge non-extremal $D_1$-$D_5$ black branes
having non-vanishing scalar curvature function on their state-space manifold $(M_4,g)$ imply an almost
everywhere weakly interacting statistical basis.

\section{Six charge non-extremal configurations}

In this section, we shall consider state-space configuration for the six parameter non-extremal
string theory black holes and focus our attention to analyze concerned state-space pair correlation
functions and present stability analysis in detail. In order to do so, we extrapolate the expression
of the entropy of four charge non-extremal $D_1$-$D_5$ solution to a non-large charge domain,
where we are no longer close to an ensemble of supersymmetric states. It is known that the leading
order entropy \cite{9603109v1} which includes all such special extremal and near-extremal cases can
be written as a function of charges $\lbrace n_i \rbrace$ and anticharges $\lbrace m_i \rbrace$ to be
\begin{eqnarray}
S(n_1,m_1,n_2,m_2,n_3,m_3):= 2 \pi (\sqrt{n_1} + \sqrt{m_1})
(\sqrt{ n_2 }+ \sqrt{m_2}) (\sqrt{n_3}+ \sqrt{m_3})
\end{eqnarray}

Incidentally, we notice from the simple brane and antibrane description that there exist an
interesting state-space interpretation which covariantly describes various statistical pair
correlation formulae arising from corresponding microscopic entropy of the aforementioned
(non) supersymmetric (non) extremal black brane configurations. Furthermore, we see for given
charges $i, j \in A_1:= \{n_1,m_1\}$; $k, l \in A_2:= \{n_2,m_2\}$; and $m,n \in A_3:= \{n_3,m_3\}$
that the intrinsic state-space pair correlations turn out to be in precise accordance with the
underlying macroscopic attractor configurations being disclosed in the special leading order
limit of the non-extremal solutions.

It is again not difficult to explore the state-space geometry of equilibrium microstates of
the six charge anticharge non-extremal black holes in $D=4$ arising from the entropy expression
emerging from the consideration of Einstein-Hilbert action. As stated earlier, we find that the
state-space Ruppeiner metric is defined by negative Hessian matrix of the non-extremal Bekenstein
-Hawking entropy with respect to the extensive variables. These variables in this case are in turn
the conserved charges-anticharges carried by the non-extremal black hole. Explicitly, we obtain that
the components of covariant state-space metric tensor over generic non-large charge domains are
\begin{eqnarray}
g_{n_1 n_1}&=& \frac{\pi}{2n_1^{3/2}} ( \sqrt{n_2} + \sqrt{m_2} )( \sqrt{n_3}+ \sqrt{m_3} ), \ 
g_{n_1 m_1}= 0 \nonumber \\
g_{n_1 n_2}&=& -\frac{\pi}{2 \sqrt{n_1 n_2}} ( \sqrt{n_3}+ \sqrt{m_3} ), \ 
g_{n_1 m_2}= -\frac{\pi}{2 \sqrt{n_1 m_2}} ( \sqrt{n_3}+ \sqrt{m_3} ) \nonumber \\
g_{n_1 n_3}&=& -\frac{\pi}{2 \sqrt{n_1 n_3}} ( \sqrt{n_2}+ \sqrt{m_2} ), \ 
g_{n_1 m_3}= -\frac{\pi}{2 \sqrt{n_1 m_3}} ( \sqrt{n_2}+ \sqrt{m_2} ) \nonumber \\
g_{m_1 m_1}&=& \frac{\pi}{2m_1^{3/2}} ( \sqrt{n_2} + \sqrt{m_2} )( \sqrt{n_3}+ \sqrt{m_3} ), \ 
g_{m_1 n_2}= -\frac{\pi}{2 \sqrt{m_1 n_2}} ( \sqrt{n_3}+ \sqrt{m_3} ) \nonumber \\
g_{m_1 m_2}&=& -\frac{\pi}{2 \sqrt{m_1 m_2}} ( \sqrt{n_3}+ \sqrt{m_3} ), \ 
g_{m_1 n_3}= -\frac{\pi}{2 \sqrt{m_1 n_3}} ( \sqrt{n_2}+ \sqrt{m_2} ) \nonumber \\
g_{m_1 m_3}&=& -\frac{\pi}{2 \sqrt{m_1 m_3}} ( \sqrt{n_2}+ \sqrt{m_2} ), \ 
g_{n_2 n_2}= \frac{\pi}{2n_2^{3/2}} ( \sqrt{n_1} + \sqrt{m_1} )( \sqrt{n_3}+ \sqrt{m_3} ) \nonumber \\
g_{n_2 m_2}&=& 0, \ 
g_{n_2 n_3}= -\frac{\pi}{2 \sqrt{n_2 n_3}} ( \sqrt{n_1}+ \sqrt{m_1} ) \nonumber \\
g_{n_2 m_3}&=& -\frac{\pi}{2 \sqrt{n_2 m_3}} ( \sqrt{n_1}+ \sqrt{m_1} ), \ 
g_{m_2 m_2}= \frac{\pi}{2m_2^{3/2}} ( \sqrt{n_1} + \sqrt{m_1} )( \sqrt{n_3}+ \sqrt{m_3} ) \nonumber \\
g_{m_2 n_3}&=& -\frac{\pi}{2 \sqrt{m_2 n_3}} ( \sqrt{n_1}+ \sqrt{m_1} ), \ 
g_{m_2 m_3}= -\frac{\pi}{2 \sqrt{m_2 m_3}} ( \sqrt{n_1}+ \sqrt{m_1} ) \nonumber \\
g_{n_3 n_3}&=& \frac{\pi}{2n_3^{3/2}} ( \sqrt{n_1} + \sqrt{m_1} )( \sqrt{n_2}+ \sqrt{m_2} ), \ 
g_{n_3 m_3}= 0 \nonumber \\
g_{m_3 m_3}&=& \frac{\pi}{2m_3^{3/2}} ( \sqrt{n_1} + \sqrt{m_1} )( \sqrt{n_2}+ \sqrt{m_2} )
\end{eqnarray}

In the entropy representation, we thus see for the non-vanishing entropy that the Hessian matrix
of entropy illustrates the nature of possible Gaussian state-space correlations between the set
of space-time parameters which in this case are nothing other than the charges on the brane and
anti-branes, if non-extremality is violated in general. Substantially, we articulate for given
non-zero value of large charges and anti charges $\{n_i, m_1 \mid i= 1,2,3 \}$ that the non-vanishing
principle component of underlying intrinsic state-space metric tensor are positive definite quantities.
It is in fact not difficult to see for distinct $ i, j, k \in \lbrace 1,2,3 \rbrace $ that the component
involving brane-brane state-space correlations $g_{n_in_i}$ and antibrane-antibrane state-space correlations
$g_{m_im_i}$ satisfy
\begin{eqnarray}
g_{n_in_i}&>& 0 \ \forall \ finite \ n_i, i=1,2,3 \nonumber \\
g_{m_im_i}&>& 0 \ \forall \ finite \ m_i, i=1,2,3
\end{eqnarray}

Furthermore, it has been observed that the ratios of diagonal components vary inversely with a multiple
of well-defined factor in the underlying parameters which change under the Gaussian fluctuations, whereas the
ratios involving off diagonal components in effect uniquely inversely vary in the of parameters of chosen set
$A_i$ of equilibrium black brane configurations. This suggests that the diagonal components weaken in relatively
controlled fashion into an equilibrium, than the off diagonal components which vary over the domain of associated
parameters defining the $D_1$-$D_5$-$P$- non-extremal non-large charge configurations. Importantly, we can easily
substantiate for the distinct $n_i, m_i \mid i \in \lbrace 1,2,3 \rbrace $ describing six charge string theory
black holes that the relative pair correlation functions have three type of relative correlation functions.
In particular, we firstly see for $i, j \in \{n_1, m_1 \}$, and $k, l \in \{n_2, m_2\}$ that the relative
correlation functions satisfy following list of scaling relations
\begin{eqnarray}
\frac{g_{ii}}{g_{jj}}&=& (\frac{j}{i})^{3/2}, \ 
\frac{g_{ii}}{g_{kk}}= (\frac{k}{i})^{3/2} (\frac{\sqrt{n_2}+\sqrt{m_2}}{\sqrt{n_3}+\sqrt{m_3}}), \ 
\frac{g_{ij}}{g_{ii}}= 0 \nonumber \\
\frac{g_{ii}}{g_{ik}}&=& -\frac{\sqrt{k}}{i} (\sqrt{n_2}+\sqrt{m_2}), \ 
\frac{g_{ik}}{g_{jk}}=  \sqrt{\frac{j}{i}}, \ 
\frac{g_{ii}}{g_{jk}}=  -\frac{\sqrt{jk}}{i^{3/2}} (\sqrt{n_2}+\sqrt{m_2}) \nonumber \\
\frac{g_{kk}}{g_{ik}}&=& -\frac{\sqrt{i}}{k} (\sqrt{n_2}+\sqrt{m_2}), \ 
\frac{g_{ij}}{g_{ik}}= 0, \ 
\frac{g_{ij}}{g_{kk}}= 0
\end{eqnarray}

The other concerned relative correlation functions are
\begin{eqnarray}
\frac{g_{ik}}{g_{il}}&=& \sqrt{\frac{l}{k}}, \ 
\frac{g_{ik}}{g_{jl}}= \sqrt{\frac{jl}{ik}}, \ 
\frac{g_{ij}}{g_{kl}}= n.d. \nonumber \\
\frac{g_{kl}}{g_{ii}}&=& 0, \ 
\frac{g_{kk}}{g_{ll}}= (\frac{l}{k})^{3/2}, \ 
\frac{g_{kl}}{g_{kk}}= 0
\end{eqnarray}

For $k, l \in \{n_2,m_2\}$, and $m, n \in \{n_3,m_3\}$, we have
\begin{eqnarray}
\frac{g_{kk}}{g_{mm}}&=& (\frac{m}{k})^{3/2} (\frac{\sqrt{n_3}+\sqrt{m_3}}{\sqrt{n_2}+\sqrt{m_2}}), \ 
\frac{g_{kl}}{g_{kk}}= 0, \ 
\frac{g_{kk}}{g_{km}}= -\frac{\sqrt{m}}{k} (\sqrt{n_3}+\sqrt{m_3})   \nonumber \\
\frac{g_{km}}{g_{lm}}&=& \sqrt{\frac{l}{k}}, \ 
\frac{g_{kk}}{g_{lm}}= -\frac{\sqrt{lm}}{k^{3/2}}(\sqrt{n_3}+\sqrt{m_3}), \ 
\frac{g_{mm}}{g_{km}}= -\frac{\sqrt{k}}{m} (\sqrt{n_2}+\sqrt{m_2}) \nonumber \\
\frac{g_{kl}}{g_{km}}&=& 0, \ 
\frac{g_{kl}}{g_{mm}}= 0
\end{eqnarray}

The other concerned relative correlation functions are
\begin{eqnarray}
\frac{g_{km}}{g_{kn}}&=& \sqrt{\frac{n}{m}}, \ 
\frac{g_{km}}{g_{ln}}= \sqrt{\frac{ln}{km}}, \ 
\frac{g_{kl}}{g_{mn}}= n. d. \nonumber \\
\frac{g_{mn}}{g_{kk}}&=& 0, \ 
\frac{g_{mm}}{g_{nn}}= (\frac{n}{m})^{3/2} , \ 
\frac{g_{mn}}{g_{mm}}= 0
\end{eqnarray}

While for $i, j \in \{n_1, m_1 \}$, and $m, n \in \{n_3,m_3\}$, we have
\begin{eqnarray}
\frac{g_{ii}}{g_{mm}}&=& (\frac{m}{i})^{3/2} (\frac{\sqrt{n_3}+\sqrt{m_3}}{\sqrt{n_1}+\sqrt{m_1}}), \ 
\frac{g_{ij}}{g_{ii}}= 0, \ 
\frac{g_{ii}}{g_{im}}= -\frac{\sqrt{m}}{i} (\sqrt{n_3}+\sqrt{m_3})  \nonumber \\
\frac{g_{im}}{g_{jm}}&=& \sqrt{\frac{j}{i}} , \ 
\frac{g_{ii}}{g_{jm}}= -\frac{\sqrt{jm}}{i^{3/2}}(\sqrt{n_3}+\sqrt{m_3}), \ 
\frac{g_{mm}}{g_{im}}= -\frac{\sqrt{i}}{m} (\sqrt{n_1}+\sqrt{m_1}) \nonumber \\
\frac{g_{ij}}{g_{im}}&=& 0, \ 
\frac{g_{ij}}{g_{mm}}= 0, \ 
\frac{g_{im}}{g_{in}}= \sqrt{\frac{n}{m}} \nonumber \\
\frac{g_{im}}{g_{jn}}&=& \sqrt{\frac{jn}{im}}, \ 
\frac{g_{ij}}{g_{mn}}= n. d., \ 
\frac{g_{mn}}{g_{ii}}= 0, \ 
\frac{g_{mn}}{g_{mm}}= 0
\end{eqnarray}

For given $i, j \in A_1:= \{n_1,m_1\}$; $k, l \in A_2:= \{n_2,m_2\}\}$;
and $m,n \in A_3:= \{n_3,m_3\}$, we thus see by utilizing $g_{n_1 m_1}=0$,
$g_{n_2 m_2}=0$ and $g_{n_3 m_3}=0$ that there are seven non-trivial relative
correlation functions for each set $A_i$, where $i= 1,2,3$, and one non-trivial
ratio in each chosen family $A_i$. It worth to mention that the scaling relations
remain similar to those obtained in the previous case, except (i) the number of
relative correlation functions has been increased, and (ii) the set of cross ratios,
\textit{viz.}, $\{ \frac{g_{ij}}{g_{kl}},\frac{g_{kl}}{g_{mn}},\frac{g_{ij}}{g_{mn}}\}$
being zero in the previous case become ill-defined for the six charge state-space configuration.
Inspecting specific pair of distinct charge sets $A_i$ and $A_j$, one finds in this case
that there are thus 24 type of non-trivial relative correlation functions.

Specifically, we see for three brane and three antibrane solutions that the ratios involving
diagonal components in the numerator with non-diagonal components in the denominator vanishes
identically $ \forall i, j, k \in \{n_1, m_1, n_2, m_2, n_3, m_3\}$. Alternatively, we thereby
appraise in this case that the set of principle components denominator ratios computed from
above state-space metric tensor reduce to
\begin{eqnarray}
\frac{g_{ij}}{g_{kk}}&=& 0 \ \forall \ i, \ j, \ k \in \{n_1, m_1, n_2, m_2, n_3, m_3\}
\end{eqnarray}

In particular for given $i, j \in A_1:= \{n_1,m_1\}$; $k, l \in A_2:= \{n_2,m_2\}\}$;
and $m,n \in A_3:= \{n_3,m_3\}$, we confirm the above fact by utilizing $g_{n_1 m_1}=0$,
$g_{n_2 m_2}=0$ and $g_{n_3 m_3}=0$ that there are total 15 type of trivial relative correlation
functions. It is not difficult to see there are five such trivial ratios in each chosen family
$\{ A_i \mid \ i= 1,2,3\}$. It worth to mention for each set $A_i$ that the trivial ratio reduce
the scaling relations which are nevertheless similar to those realized in the previous case,
except the fact that the number of relative correlation functions has been ill-defined.
Inspecting a pair of distinct charge sets $A_i$ and $A_j$, one finds in this case that there
is an unique kind of ill-defined relative correlations, and thus there are in total three type
of divergent relative correlation functions.

As noticed in the previous configuration, it is not difficult to analyze the local stability
for the higher charged string theory non-extremal black holes, as well. In particular, one can
easily determine the principle minors associated with the state-space metric tensor and thus we
argue that all the principle minors must be positive definite. In this case, we may adroitly compute
the principle minors from the Hessian matrix of associated entropy concerning the three charge and
three anticharged black holes. In fact, after some simple manipulations we discover that the local
stability criteria on the lower dimensional hyper-surfaces and two dimensional surface of underlying
state-space manifold are respectively given by the following relations
\begin{eqnarray}
p_1 &=& \frac{\pi}{2n_1^{3/2}} ( \sqrt{n_2} + \sqrt{m_2} )( \sqrt{n_3}+ \sqrt{m_3} ) \nonumber \\
p_2 &=& \frac{1}{4} \frac{\pi^2}{(n_1m_1)^{3/2}}(\sqrt{n_2}+\sqrt{m_2})^2(\sqrt{n_3}+\sqrt{m_3})^2 \nonumber \\
p_3 &=&  \frac{1}{8} \frac{\pi^3}{(n_1m_1n_2)^{3/2}} \sqrt{m_2} (\sqrt{n_3}+\sqrt{m_3})^3 (\sqrt{n_2}+\sqrt{m_2})
 (\sqrt{n_1}+\sqrt{m_1}) \nonumber \\
p_4 &=& 0 
\end{eqnarray}

For all physically admitted values of associated charges and anticharges of the non-extremal
string theory black holes, we thus ascertain that the minor constraint, {\it viz.}, $p_2>0$
inhibits the domain of assigned brane anti-brane charges that it must be a positive definite
real number, while the constraint $p_3>0$ imposes that the charges must respectively satisfy
desired state-space minor conditions. In particular, these constraints enables us to investigate
the nature of the state-space geometry of string theory black holes. We have further observed
that the presence of planer and hyper planer instabilities exist for the  non-extremal black
holes. It is worth to mention that the $p_4(n_i, m_i)=0$ exhibits that the four dimensional
state-space configurations are not stable for any value of the brane and antibrane numbers.
This altogether demand for definite restriction on the allowed value of the parameters.

Similarly we find that the principle minor $p_5$ remains non-vanishing for all values of charges
on the constituent brane and anti branes. The generic expression of the minor $p_5$ may further be
easily computed from the general minor formula \cite{RotBHs}. An explicit calculation specifically
finds that the hyper-surface minor $p_5$ take fairly non-trivial value in general. However, the
simplest values of the brane and antibrane charges that they be identical implies that the minor
$p_5$ reduces to the specific value of
\begin{eqnarray}
p_5(k)= -64 \frac{\pi^5}{k^{5/2}}
\end{eqnarray}

Thus for the identical values of the brane antibrane charges, the minor $p_5<0$ respectively implies
that the non-extremal black hole solutions under consideration are not stable over the possible choice
of the state-space configurations. In order to obtain the highest minor $p_6$, we in general need to
compute determinant of the metric tensor which finally reduces as the function of the charges on branes
and antibranes. Moreover, it is not difficult to demonstrate the global stability on the full state-space
configuration, which may in fact be carried forward by computing determinant of the state-space metric tensor.
In this case, one observes that the exact expression of the determinant of the intrinsic state-space metric tensor is
\begin{eqnarray}
\Vert g \Vert&=& -\frac{ \pi^6 }{16} (n_1m_1n_2m_2n_3m_3)^{-3/2} (\sqrt{n_2} +
\sqrt{m_2} )^2 (\sqrt{n_3} + \sqrt{m_3} )^3 (\sqrt{n_1} + \sqrt{m_1} )^3 \nonumber \\&&
( n_2 \sqrt{m_1 n_3} + n_2 \sqrt{m_1m_3} + 2 \sqrt{n_2 m_1  m_2 n_3} + 2 \sqrt{n_2 m_1 m_2 m_3}
 + m_2 \sqrt{m_1 n_3}  \nonumber \\ &&+  m_2 \sqrt{m_1 m_3} + n_2 \sqrt{n_1  n_3} + n_2
\sqrt{n_1  m_3} +  2 \sqrt{n_1 n_2 m_2 n_3} + 2 \sqrt{ n_1 n_2 m_2
m_3 } \nonumber \\&&+ m_2 \sqrt{n_1  n_3} + m_2 \sqrt{n_1  m_3} )
\end{eqnarray}

which in turn never vanishes for domain of given non-zero brane antibrane charges, except for following
state-space extreme values of the charges, when the brane and antibrane charges $n_i$, $m_i$ belong to
\begin{eqnarray}
B&:=&\lbrace \ (n_1,n_2,n_3,m_1,m_2,m_3) \mid  n_2 \sqrt{m_1 n_3}+ n_2
\sqrt{m_1 m_3}+ 2 \sqrt{n_2 m_1  m_2 n_3}+ \nonumber \\&& 2\sqrt{n_2 m_1  m_2m_3}+
m_2 \sqrt{m_1 n_3}+  m_2 \sqrt{m_1 m_3}+ n_2 \sqrt{n_1  n_3}
+ n_2 \sqrt{n_1 m_3}  \nonumber \\&&+  2 \sqrt{n_1 n_2 m_2  n_3} + 2 \sqrt{ n_1
n_2 m_2 m_3 }+ m_2 \sqrt{n_1  n_3} + m_2 \sqrt{n_1  m_3} =\ 0 \ \rbrace
\end{eqnarray}

We may further note that the entire state-space configuration remains positive definite for potential
value of the $n_i,m_i$. We thus observe that the underlying state-space geometry of six charge non-extremal
string theory configurations are in well compliance and in turn they generically correspond to a non-degenerate
fluctuating statistical basis as an intrinsic Riemannian manifold $N:= M_6\setminus B$. Furthermore, we see that
the components of the covariant Riemann tensors may become zero for definite values of the charges on branes and
antibranes. In addition, the Ricci scalar curvature diverges at the same set of points on state-space manifold
$(M_6,g)$, as that of the roots of the determinant of metric tensor, \textit{viz.}, the points defined by the set $B$.

There exists an akin single higher degree polynomial equation on which we precisely find that the Ricci scalar
curvature becomes null, and exactly at these points defining the state-space configuration of the underlying
(extremal or near-extremal or general) non-large charge black hole system, at which it corresponds to some
non-interacting statistical system, where the state-space manifold $(M_6,g)$ is curvature free. A systematic
calculation further shows that the general expression for the Ricci scalar is quite involved, and even for
equal brane charges $n_1:=n$; $n_2:=n$; $n_3:=n$; and  equal antibrane charges $m_1:=m$; $m_2:=m$; $m_3:=m$
the result do not sufficiently simplifies. Nevertheless, we find for the identical large values of brane and
antibrane charges $n:=k$ and $m:=k$ \cite{BNTBull} that there exist an attractive state-space configuration
for which the expression of corresponding curvature scalar reduces to a particular small negative value of
\begin{eqnarray}
R(k)= -\frac{15}{16} \frac{1}{\pi k^{3/2}}
\end{eqnarray}

\section{Multi-centered $D_6D_4D_2D_0$ Black Branes}

The present section explores the state-space manifold containing the both single
centered black brane solutions and multi-centered black brane configurations, \textit{viz.},
we shall study the state-space geometry whose co-ordinates are defined in terms of four charges
of the $D_6D_4D_2D_0$ black brane configurations. Here, we shall explicitly present the analysis
of the state-space correlations arising from the entropy of stationary single-centered systems
as well as that the double centered black hole molecule configurations. Such multi-centered
black hole configurations have recently been examined by the so called pin-sized $D$-brane
systems \cite{0705.2564v1, 0702146v2} and thus we intend to realize underlying state-space
geometry arising from the counting entropy of the number of microstates of zoo of entropically
dominant multi-centered black brane configurations along with usual single centered black holes.

It has been shown \cite{0705.2564v1, 0702146v2} in suitable parameter regimes that the
multi-centered entropy dominates the single centered entropy in the uniform large charge limit.
Following \cite{BNTBull}, we shall here investigate the state-space geometric implication for the
single center and two centers of the multi-centered $D_6D_4D_2D_0$ systems. In this connection,
we may consider a charge $\Gamma= \sum_i \Gamma_i$ obtained by wrapping the $D_4$, $D_2$ and $D_0$
branes around various cycles of a compact space $X$, and the concerned charges are scaled as
$\Gamma \rightarrow \Lambda \Gamma$, and then there exist two centered brane solution with horizon
entropy scaling as $\Lambda^3$; while that of the single centered entropy simply scales as $\Lambda^2$.
More properly as analyzed in \cite{0705.2564v1, 0702146v2}, consider the type IIA string theory compactified
on a product of three two-tori $X:= T_1^2 \times T_2^2 \times T_3^2$. Then, the entropy as a function
the charge $\Gamma $ corresponding to $ p0 $ $ D_6$ branes on $ X$, $ p $ $D_4$ branes on
$(T_1^2 \times T_2^2)+ (T_2^2 \times T_3^2)+ (T_3^2 \times T_1^2)$, $q$ $ D_2$ branes on
$(T_1^2 + T_2^2 + T_3^2)$ and $q0$ $D_{0}$ branes is given by
\begin{eqnarray}
S(\Gamma):= \pi \sqrt{-4 p^3 q0+ 3 p^2 q^2+ 6 p0 pqq0- 4 p0q^3-(p0q0)^2}
\end{eqnarray}

The state-apace geometry constructed out of the equilibrium state of the four charged
$D_6D_4D_2D_0$ black branes resulting from the entropy may thus be easily computed as
earlier from the negative Hessian matrix of the entropy with respect to the $D_6$, $D_4$,
$D_2$, $D_0$ brane charges $\Gamma_i:= (p_i^\Lambda, q_{\Lambda,i})$ which in effect form
the coordinates of the intrinsic state-space manifold. Explicitly, we find that the components
of the covariant metric tensor are given as
\begin{eqnarray}
g_{p0p0}&=& -4 \pi \frac{-3p^2q^2q0^2+ 3pq^4q0- q^6+ p^3q0^3}
{(-4 p^3 q0+ 3 p^2 q^2+ 6 p0 pqq0- 4 p0q^3-(p0q0)^2)^{3/2}} \nonumber \\
g_{p0p}&=& 6 \pi \frac{-p^3q0^2q+ 2 p^2q0q^3+ p^2q0^3p0- pq^5- 2pq^2p0q0^2+ p0q^4q0}
{(-4 p^3 q0+ 3 p^2 q^2+ 6 p0 pqq0- 4 p0q^3-(p0q0)^2)^{3/2}} \nonumber \\
g_{p0q}&=& -12 \pi \frac{2p^3q^2q0+ p^2qq0^2p0- 2pq^3q0p0- q^4p^2+ q^5p0- p^4q0^2}
{(-4 p^3 q0+ 3 p^2 q^2+ 6 p0 p q q0- 4 p0q^3-(p0q0)^2)^{3/2}} \nonumber \\
g_{p0q0}&=& -\pi \frac{-6p^4qq0+ 3p^2q^2q0p0- 9pqq0^2 p0^2+ 5q^3 p^3-
6q^4 p0 p+ 6q^3 p0^2 q0+ 6p0 q0^2 p^3+ p0^3q0^3}
{(-4 p^3 q0+ 3 p^2 q^2+ 6 p0 p q q0- 4 p0q^{3}-(p0q0)^2)^{3/2}} \nonumber \\
g_{pp}&=& -12 \pi \frac{p^4q0^2- p^3q^2q0- 3p^2qq0^2p0+ 4pq^3q0p0- p0^2q0^2q^2+ p0^2q0^3p- q^5p0}
{(-4 p^3 q0+ 3 p^2 q^2+ 6 p0 p q q0- 4 p0q^3-(p0q0)^2)^{3/2}} \nonumber \\
g_{pq}&=& 3 \pi \frac{2p^4qq0- 2p0q0^2p^3+ 3p^2q^2q0p0- 3q^3p^3+ 2q^4p0p- pqq0^2p0^2-
2q^3p0^2q0+ (p0q0)^3 }{(-4 p^3 q0+ 3 p^2 q^2+ 6 p0 p q q0- 4 p0q^3-(p0q0)^2)^{3/2}} \nonumber \\
g_{pq0}&=& - 12 \pi \frac{p^5q0- 2p^3q0p0q- p^4q^2+ 2p^2q^3p0+ pq^2p0^2q0- p0^2q^4}
{(-4 p^3 q0+ 3 p^2 q^2+ 6 p0 p q q0- 4 p0q^3-(p0q0)^2)^{3/2}} \nonumber \\
g_{qq}&=& -12 \pi \frac{4p^3q0p0q- p^2q^3p0- p^2q0^2p0^2- 3pq^2p0^2q0+ p0^2q^4- p^5q0+ p0^3qq0^2}
{(-4 p^3 q0+ 3 p^2 q^2+ 6 p0 p q q0- 4 p0q^3-(p0q0)^2)^{3/2}} \nonumber \\
g_{qq0}&=& 6 \pi \frac{-p^5q+ 2p^3q^2p0- 2p^2qp0^2q0+ p0p^4q0- p0^2q^3p+ p0^3q^2q0}
{(-4 p^3 q0+ 3 p^2 q^2+ 6 p0 p q q0- 4 p0q^3-(p0q0)^2)^{3/2}} \nonumber \\
g_{q0q0}&=& -4 \pi \frac{-p^6+ 3p^4p0q- 3p0^2p^2q^2+ p0^3q^3}
{(-4 p^3 q0+ 3 p^2 q^2+ 6 p0 p q q0- 4 p0q^{3}-(p0q0)^2)^{3/2}}
\end{eqnarray}

In order to simplify the presentation, we shall define $X_a=(p0,p,q,q0)$ and subsequently use the following
set of notations $1 \leftrightarrow p_0, 2 \leftrightarrow p, 3 \leftrightarrow q, 4 \leftrightarrow q_0$.
Employing the either of above notations, we observe from the definition that the ascertained statistical pair
correlations may in turn be accounted by simple microscopic descriptions which can be expressed in terms of
the brane charges connoting an ensemble of microstates of the multicentered black hole configuration. Furthermore,
it is in fact evident that the principle components of statistical pair correlations are positive definite
for all allowed values of the concerned parameters of the $D_6$-$D_4$-$D_2$-$D_0$ black holes. As a result,
we can easily see that the concerned state-space metric constraints are defined by
\begin{eqnarray} \label{multipositivity}
g_{ii}(X_a)&>& 0 \ \forall \ i \in \{1,2,3,4\} \mid m_{ii} < 0 
\end{eqnarray}

The principle components of state-space metric tensor $\lbrace g_{ii}(X_a) \vert \ i=1,2,3,4 \rbrace$
essentially signify a set of definite heat capacities (or the related compressibility) whose positivity
apprises that the black brane solution complies an underlying locally equilibrium statistical configuration.
It is intriguing to note that the positivity of the components $g_{ii}$ require that the brane charges of
associated multicentered $D_6$-$D_4$-$D_2$-$D_0$ black holes should satisfy the above constraints. This is
indeed admissible because of the fact that the brane configuration divulges physically stable system for
all values of the brane charges satisfying  Eqn. (\ref{multipositivity}) with
\begin{eqnarray}
m_{11}&:=&- 3p^2q^2q0^2+ 3pq^4q0- q^6+ p^3q0^3 \nonumber \\
m_{22}&:=& p^4q0^2- p^3q^2q0-3p^2qq0^2p0+ 4pq^3q0p0 \nonumber \\&& - p0^2q0^2q^2+ p0^2q0^3p- q^5p0 \nonumber \\
m_{33}&:=& 4p^3q0p0q- p^2q^3p0- p^2q0^2p0^2- 3pq^2p0^2q0 \nonumber \\&& + p0^2q^4- p^5q0+ p0^3qq0^2 \nonumber \\
m_{44}&:=& -p^6+ 3p^4p0q- 3p0^2p^2q^2+ p0^3q^3
\end{eqnarray}

From the above expressions of metric tensor, we visualize that the ratios of the principle components
of statistical pair correlations vary as definite function of the asymptotic charges; while that of
the off-diagonal correlations modulate slightly differently. Interestingly, it follows for the distinct
$i,j,k,l \in \lbrace 1,2,3,4 \rbrace $ that the admissible statistical pair correlations thus connoted
are consisting of diverse scaling properties. The set of nontrivial relative correlations signifying
possible scaling relations of the state-space correlations may nicely be depicted by
\begin{eqnarray}
C_r&=& \bigg\{\frac{g_{11}}{g_{12}},\frac{g_{11}}{g_{13}},\frac{g_{11}}{g_{14}},\frac{g_{11}}{g_{22}},\frac{g_{11}}{g_{23}},
\frac{g_{11}}{g_{24}},\frac{g_{11}}{g_{33}},\frac{g_{11}}{g_{34}},\frac{g_{11}}{g_{44}}, 
\frac{g_{12}}{g_{13}},\frac{g_{12}}{g_{14}},\frac{g_{12}}{g_{22}},\frac{g_{12}}{g_{23}},\frac{g_{12}}{g_{24}},
\frac{g_{12}}{g_{33}}, \nonumber \\ && \frac{g_{12}}{g_{34}},\frac{g_{12}}{g_{44}}, 
\frac{g_{13}}{g_{14}},\frac{g_{13}}{g_{22}},\frac{g_{13}}{g_{23}},\frac{g_{13}}{g_{24}},\frac{g_{13}}{g_{33}},
\frac{g_{13}}{g_{34}},\frac{g_{13}}{g_{44}}, 
\frac{g_{14}}{g_{22}},\frac{g_{14}}{g_{23}},\frac{g_{14}}{g_{24}},\frac{g_{14}}{g_{33}},\frac{g_{14}}{g_{34}},
\frac{g_{14}}{g_{44}},\nonumber \\ && 
\frac{g_{22}}{g_{23}},\frac{g_{22}}{g_{24}},\frac{g_{22}}{g_{33}},\frac{g_{22}}{g_{34}},\frac{g_{22}}{g_{44}},
\frac{g_{23}}{g_{24}},\frac{g_{23}}{g_{33}},\frac{g_{23}}{g_{34}},\frac{g_{23}}{g_{44}},
\frac{g_{24}}{g_{33}},\frac{g_{24}}{g_{34}},\frac{g_{24}}{g_{44}}, 
\frac{g_{33}}{g_{34}},\frac{g_{33}}{g_{44}},
\frac{g_{34}}{g_{44}} \bigg \}
\end{eqnarray}

The local stability condition of the underlying statistical configuration under the Gaussian
fluctuations requires that all the principle components of the fluctuations should be positive
definite, i.e. for given set of state-space variables $ \Gamma_i:= (p_i^\Lambda, q_{\Lambda,i})$
one must demands that $\lbrace g_{ii}(\Gamma_i) > 0; \ \forall i= 1, 2\rbrace$. In particular,
it is important to note that this condition is not sufficient to insure the global stability
of chosen multicentered configuration and thus one may only accomplish certain locally equilibrium
statistical configuration. It is however worth to mention that the complete stability condition
requires that all the principle components of the Gaussian fluctuations should be positive definite
and the others components of the fluctuations should vanish. In order to ensure this condition,
we can observe that all the principle components and all the principle minors of the metric tensor
must be strictly positive definite. This implies that the global stability condition constraint
the allowed domain of the parameters of black hole configurations, which are interestingly expressed
by the following set of simultaneous equations
\begin{eqnarray}
p_1&=& -4 \pi \frac{(-3 p^2 q^2 q0^2+ 3 p q^4 q0- q^6+ q0^3 p^3)}
{(-4 p^3 q0+ 3 p^2 q^2+ 6 p0 p q q0- 4 p0 q^3- p0^2 q0^2)^{-3/2}} \nonumber \\
p_2&=& -12 \pi^2 \frac{(q0^4 p^4- 4 q^2 q0^3 p^3+ 6 q^4 q0^2 p^2- 4 q^6 q0 p + q^8)}
{(4 p^3 q0- 3 p^2 q^2- 6 p0 p q q0 + 4 p0 q^3 + p0^2 q0^2)^{-2}}  \nonumber \\
p_3&=& -36 \pi^3 \frac{(-3 p^2 q^2 q0^2+ 3 p q^4 q0- q^6+ q0^3 p^3)}
{(-4 p^3 q0+ 3 p^2 q^2+ 6 p0 p q q0- 4 p0 q^3- p0^2 q0^2)^{-3/2}}  
\end{eqnarray}

In addition, it is evident that the local stability of the full state-space configuration
can likewise be determined by computing the determinant of the metric tensor of concerned
state-space geometry. Here, we may easily compute a compact formula for the determinant of
the metric tensor as the function of various possible values of brane charges, and in particular,
our intrinsic geometric analysis assigns the following constant expression to the determinant
of the metric tensor
\begin{eqnarray}
\Vert g \Vert= 9 \pi^4
\end{eqnarray}

As the determinant of basic state-space metric tensor is constant and positive quantity in
the viewpoints of large charge consideration in which one acquires a non-vanishing central
charge of corresponding $D_6$-$D_4$-$D_2$-$D_0$ CFT configurations \cite{0705.2564v1, 0702146v2}.
Our analysis herewith discovers that there exists non-degenerate state-space geometry for the
leading multi-centered configurations. Furthermore, it is worth to note that the determinant of
the metric tensor takes positive definite form, which in turn shows that there is positive definite
volume form on the concerned state-space manifold $(M_4,g)$ of the multi-centered $D_6$-$D_4$-$D_2$-$D_0$
black brane configurations at the leading order contributions.

Intelligibly, it further follows from the fact that the responsible equilibrium entropy tends to
its maximum value, while the same culmination may not remain valid on the chosen planes or
hyper-planes of the entire state-space manifolds of the single centered and double centered
configurations. It is thus envisaged for the either single or double center descriptions or
the dual CFT descriptions that the multi-centered black branes do correspond to intrinsically
stable statistical configurations. Thus, it is indeed plausible that the underlying ensemble
of CFT microstates upon subleading higher derivative corrections live in the same basis of
$D_6$-$D_4$-$D_2$-$D_0$ brane charges.

\subsection{State-space correlations of the single center configurations}

For the charges, $ p0:=0$; $p:=6 \Lambda$; $q:=0$; $q0:=-12 \Lambda$; describing
the single center configurations considered in \cite{0705.2564v1, 0702146v2},
we see that the above state-space correlation functions reduce to the
following values
\begin{eqnarray}
g_{11}&=& \pi \sqrt{2}, \ 
g_{13}= \frac{3}{2} \pi \sqrt{2} 
=-g_{22}  \nonumber\\
g_{24}&=& \frac{3}{4} \pi \sqrt{2} 
= -g_{33}, \ 
g_{44}= \frac{1}{8} \pi \sqrt{2} \nonumber\\
g_{12}&=& 0 
=g_{14}
=g_{23}
=g_{34}
\end{eqnarray}

Following the previously acclaimed notations, we observes that the statistical pair
correlations being accounted by simple state-space characterization can be expressed
in terms of the brane charges. Furthermore, an easy analysis find that all the principle
components of the statistical pair correlations are positive definite for chosen value
of parameters of the single center black holes. In particular, we see for $ p0:=0$;
$p:=6 \Lambda$; $q:=0$; $q0:=-12 \Lambda$ that the concerned state-space metric
constraints can for all $\Lambda$ be depicted as
\begin{eqnarray}
g_{ii}(X_a)&>& 0 \ \forall \ i=1,3 \nonumber \\
g_{jj}(X_a)&<& 0 \ \forall \ j=2,4 
\end{eqnarray}

The principle components of state-space metric tensor $\lbrace g_{ii} \vert \ i=1,2,3,4 \rbrace$
signifying a set of heat capacities (or the related compressibility) do not all find positive values.
Here, a violation of the positivity of heat capacity apprises that the corresponding single center
black brane solution corresponds to a locally unstable statistical configuration over the Gaussian
fluctuations. It is thus important to mention that the positivity of principle components do not holds
for the above set of brane charges associated with the single centered $D_6D_4D_2D_0$ black branes.

In analyzing the other state-space constraints. we see that the relative correlations defined
as $c_{ijkl}:= g_{ij}/g_{kl}$ reduce to the three set of constant values. Firstly following the proclaimed
procedure, we find that there are only 15 non vanishing finite ratios defining the relative state-space
correlation functions for the single center configuration
\begin{eqnarray}
c_{1113}&=& \frac{2}{3}= -c_{1122}, \ 
c_{1124}= \frac{4}{3}= -c_{1133} \nonumber\\
c_{1144}&=& 8, \ 
c_{1322}= -1= c_{2433} \nonumber\\
c_{1324}&=& 2= -c_{1333}, \ 
c_{2233}= 2= -c_{2224} \nonumber\\
c_{1344}&=& 12= -c_{2244}, \ 
c_{2444}= 6= -c_{3344}
\end{eqnarray}


Furtheremore, an observation finds that the set of vanishing ratios of relative correlation functions is
\begin{eqnarray}
C_R^0&:=& \{ c_{1213}, c_{1224}, c_{1222}, c_{1224}, c_{1233}, c_{1244}, c_{1422}, \nonumber\\ &&
c_{1424}, c_{1433}, c_{1444}, c_{2324}, c_{2333}, c_{2344}, c_{3444} \} \nonumber\\
&=& \{ 0 \}
\end{eqnarray}


In particular, we notice for $ p0:=0$; $p:=6 \Lambda$; $q:=0$; $q0:=-12 \Lambda$
that there exist limiting ill-defined relative correlations. In particular,
the concerned ratios get numeric exception and they receive a division by zero
when approaching the single centered configuration. Thus, the  characterization
of the relative state-space pair correlation may be accomplished by the set
\begin{eqnarray}
C_R^{\infty}&:=& \{c_{1112}, c_{1114}, c_{1123}, c_{1134}, c_{1223}, c_{1234}, c_{1314}, c_{1323}, \nonumber\\ &&
c_{1334}, c_{1423}, c_{1434}, c_{2223}, c_{2234}, c_{2334}, c_{2434}, c_{3334} \} \nonumber\\ &=& \ \{ \infty \}
\end{eqnarray}


\subsection*{State-space stability of single center $D_6$-$D_4$-$D_2$-$D_0$ configurations}
Furthermore, we see that the entropy corresponding to single center specification takes
to a constant value of $ S( \Gamma= \Lambda (0,6,0,-12))= \pi \sqrt{10368} \Lambda^2 $.
Whilst, it is interesting to note that the possible stability of internal configurations
under the Gaussian fluctuations reduce to the positivity of
\begin{eqnarray}
p_1= \sqrt{2} \pi, \ 
p_2= -3 \pi^2, \ 
p_3= 9 \sqrt{2} \pi^3
\end{eqnarray}

We thus find for the chosen values of brane charges which physically describe single
center system that the statistical configuration has definite stability and instability
character. In particular, positivity of $p_1(0,6 \Lambda,0,-12 \Lambda)$ shows that the
underlying brane configurations are locally stable on an intrinsic state-space line.
Nevertheless, we observe for the chosen value of brane charges that the two dimensional
surfaces of the single center state-space configurations are not stable. This han in turn
been easily ascertained via the fact that the associated surface minor constraint is not satisfied.
Specifically, it turns out for chosen set of charge $(0,6 \Lambda,0,-12 \Lambda)$ that the
system exhibits a negative surface minor, {\it viz.}, we have $p_2(0,6 \Lambda,0,-12 \Lambda)<0$.
Similarly, one may however notice that the system turns out to be stable on three dimensional
hyper-surfaces against the single center configurations. The argument follows directly from the fact
that the hypersurface minor $p_3(0,6 \Lambda,0,-12 \Lambda)$ picks up a positive definite value.

More generally, it interesting to note that the general expression of the determinant of the
metric tensor defined as $g = 9 \pi^4$ remains constant for entire domain of brane parameters.
In addition, we find for all non zero entropy solution that the state-space scalar curvature
signifying global correlation volume of an underlying statistical system has no divergence.
As expected further from \cite{BNTBull} that the leading order entropy solutions defining
the single solutions confirm for all admissible values of brane charges that their state-space
correlation volume vary as an inverse function of the single center brane entropy arising from
the degeneracy of equilibrium statistical configurations.

\subsection{State-space correlations of the two center configurations}

In this subsection, we shall explicitely present the state-space geometry of $D_6$-$D_4$-$D_2$-$D_0$
black holes in string theory carrying set of repective $D$-brane charges. We notice that these solutions
carry different state-space pair correlations. For given $\Lambda$, this follows from the fact that the
two center black holes in general have particular correlations which are not the same for both of the
centers or as that of the single center counter parts. We nevertheless find that the global state-space
correlations which characterize stability of the vacuum string theory configurations of either center
do not change for the choice of brane charges considered in \cite{0705.2564v1, 0702146v2}.

\subsubsection{State-space correlations at the first center}

For the value of beane charges $p0:=1$; $p:=3\Lambda$; $q:=6\Lambda^2$; and $q0:=-6\Lambda$
defining first center of the two center $D_6$-$D_4$-$D_2$-$D_0$ configurations, we have
the following components of the state-space metric tensor
\begin{eqnarray} \label{gij1}
g_{11}&=& 108 \pi \Lambda^3 \frac{6 \Lambda^2+ 12 \Lambda^4+ 8 \Lambda^6+ 1}{( 3 \Lambda^4- 1 )^{3/2}}, \ 
g_{12}= -54 \pi \Lambda^2 \frac{7 \Lambda^2+ 16 \Lambda^4+ 12 \Lambda^6+ 1}{( 3 \Lambda^4- 1 )^{3/2}} \nonumber\\
g_{13}&=& 54 \pi \Lambda^3 \frac{4 \Lambda^2+ 4 \Lambda^4+ 1}{( 3 \Lambda^4- 1 )^{3/2}}, \ 
g_{14}= - \pi \frac{18 \Lambda^4+ 27 \Lambda^6- 1}{( 3 \Lambda^4- 1 )^{3/2}} \nonumber\\
g_{22}&=& 18 \pi \Lambda \frac{13 \Lambda^2+ 30 \Lambda^4+ 24 \Lambda^6+ 2}{( 3 \Lambda^4- 1 )^{3/2}}, \ 
g_{23}= -3 \pi \frac{42 \Lambda^4 + 12 \Lambda^2 + 45 \Lambda^6 + 1 }{(3 \Lambda^4-1)^{3/2}} \nonumber\\
g_{24}&=& 9 \pi \Lambda^3 \frac{1+ 2\Lambda^2}{( 3 \Lambda^4- 1 )^{3/2}}, \ 
g_{33}= 3 \pi \Lambda \frac{2+ 9\Lambda^2+ 12\Lambda^4}{( 3 \Lambda^4- 1 )^{3/2}}  \nonumber\\
g_{34}&=& -\frac{3}{2} \pi \Lambda^2 \frac{1+ 3\Lambda^2}{( 3 \Lambda^4- 1 )^{3/2}}, \ 
g_{44}= \frac{1}{2} \pi \Lambda^3 \frac{1}{( 3 \Lambda^4- 1 )^{3/2}}  
\end{eqnarray}

Simplifying subsequent notations by defining $c_{ijkl}:= g_{ij}/g_{kl}$, we then see at
the first center of the two center $D_6$-$D_4$-$D_2$-$D_0$ that the relative state-space
correlations describing concerned statistical system are physically sound in nature.
We in this case see that it is not difficult to compute the $c_{ijkl}$. Nevertheless,
the exact expression for the set of $c_{ijkl}$ is quite involved and thus we relegate
them to the Appendix (A).

\subsubsection{State-space correlations at the second center}

The value of charges $p0:=-1$; $p:=3\Lambda$; $q:=-6\Lambda^2$; and $q0:=-6\Lambda$
define the second center of the two center configurations for which we have the following
limiting values of the state-space pair correlation functions
\begin{eqnarray} \label{gij2}
g_{11}&=& 108 \pi \Lambda^3 \frac{6 \Lambda^2+ 12 \Lambda^4+ 8 \Lambda^6+ 1}{( 3 \Lambda^4- 1 )^{3/2}}, \ 
g_{12}= 54 \pi \Lambda^2 \frac{7 \Lambda^2+ 16 \Lambda^4+ 12 \Lambda^6+ 1}{( 3 \Lambda^4- 1 )^{3/2}} \nonumber\\
g_{13}&=& 54 \pi \Lambda^3 \frac{4 \Lambda^2+ 4 \Lambda^4+ 1}{( 3 \Lambda^4- 1 )^{3/2}}, \ 
g_{14}= \pi \frac{18 \Lambda^4+ 27 \Lambda^6- 1}{( 3 \Lambda^4- 1 )^{3/2}} \nonumber\\
g_{22}&=& 18 \pi \Lambda \frac{13 \Lambda^2+ 30 \Lambda^4+ 24 \Lambda^6+ 2}{( 3 \Lambda^4- 1 )^{3/2}}, \ 
g_{23}= 3 \pi \frac{42 \Lambda^4 + 12 \Lambda^2 + 45 \Lambda^6 + 1 }{( 3 \Lambda^4- 1 )^{3/2}} \nonumber\\
g_{24}&=& 9 \pi \Lambda^3 \frac{1+ 2\Lambda^2}{( 3 \Lambda^4- 1 )^{3/2}}, \ 
g_{33}= 3 \pi \Lambda \frac{2+ 9\Lambda^2+ 12\Lambda^4}{( 3 \Lambda^4- 1 )^{3/2}}  \nonumber\\
g_{34}&=& \frac{3}{2} \pi \Lambda^2 \frac{1+ 3\Lambda^2}{( 3 \Lambda^4- 1 )^{3/2}}, \ 
g_{44}= \frac{1}{2} \pi \Lambda^3 \frac{1}{( 3 \Lambda^4- 1 )^{3/2}}
\end{eqnarray}

Employing the previously defined notations, it has similarly seen that the relative
correlations $c_{ijkl}:= g_{ij}/g_{kl}$ of the state-space configuration concerning
second center of the $D_6$-$D_4$-$D_2$-$D_0$ system simplify to the one which have
been presented in the Appendix (B).

\subsection*{State-space stability of double center $D_6$-$D_4$-$D_2$-$D_0$ configurations}
We shall now consider state-space stability for the two center black brane configurations
and analyzes the related positivity properties of their underlying statistical pair correlation
functions and correlation volumes for the basins of $D_6$-$D_4$-$D_2$-$D_0$ configurations.
At the first and second centers of the two center $D_6$-$D_4$-$D_2$-$D_0$ configurations,
we find in turn that the mentioned statistical pair correlations can be simply accounted
by a common factor of the charges $\Gamma_i$. These notions further receive supports from microscopic
descriptions as well that an ensemble of microstates of the multicentered black hole configurations
could effectively be expressed in terms of $\Lambda$ as such basis are simply connoted via the
invariant brane charges $\{ \Gamma_i \}$.

As indicated by Denef and Moore in \cite{0705.2564v1, 0702146v2} that
the two centered bound state configurations arise with charge centers
$ \Gamma_1 = (1,3 \Lambda, 6 \Lambda^2 ,-6 \Lambda) $ and
$ \Gamma_2 = (-1,3 \Lambda, -6 \Lambda^2 ,-6 \Lambda) $. Thus, we shall focus
our attention for these charge centers and analyze their state-space quantities
as the function of $\Lambda $. It is apparent for some given $ \Lambda $ that
the entropies at either of the two centers $ \Gamma_1$, $\Gamma_2$ match,
and in particular, we find that the double center entropy  varies as
\begin{eqnarray}
S(\Gamma_1)=  S(\Gamma_2)= \pi \sqrt{108\Lambda^6-36\Lambda^2} \sim \Lambda^3
\end{eqnarray}

For given charge centers $\Gamma_1$ and $ \Gamma_2$, we may apart from definite scaling in $\Lambda$
appreciate over an equilibrim statistical basis that either of the state-space pair correlation function
as defined in the Eqn. (\ref{gij1}) and Eqn. (\ref{gij2}) can be realized as an even function of the
parameter $\Lambda$. Similarly, one can contemplate possible nature of the pair correlation functions
over the jump of one center to the other. In turn for chosen  $\Gamma_1$ and $ \Gamma_2$, we find that
the above two center $D_6$-$D_4$-$D_2$-$D_0$ configurations form two type of state-space pair correlation
functions. In particular, we see from the Eqn. (\ref{gij1}) and Eqn. (\ref{gij2}) that the two proclaimed
set of pair correlations are
\begin{eqnarray}
C_{ij}^{(1)}(\Gamma)&:=& \{ g_{ij}(\Gamma_1)=  g_{ij}(\Gamma_2);
(i,j) \in \{(1,1),(1,3),(2,2),(2,4),(3,3),(3,4),(4,4) \} \} \nonumber \\
C_{ij}^{(2)}(\Gamma)&:=& \{ g_{ij}(\Gamma_1)= -g_{ij}(\Gamma_2);
(i,j) \in \{(1,2),(1,4),(2,3) \} \}
\end{eqnarray}

It has explicitly been seen for non-vanishing $\Lambda$ that the state-space pair correlations belonging to
$C^{(1)}$ remain the same for both the centers, while the pair correlations belonging to the set $C^{(2)}$ change
their signature. The present analysis implies that the principle components of the metric tensor defining
equilibrium statistical pair correlations are positive definite for all allowed values of the parameter $\Lambda$.
In fact, the $\Lambda$ being the single parameter for the both first and second centers of the two center
$D_6$-$D_4$-$D_2$-$D_0$ black branes describes potential stability and state-space correlation properties
of the $D_6$-$D_4$-$D_2$-$D_0$ multi-center configurations. As a result, we see for all $\Lambda$ and for
either of the two centers, \textit{viz.}, $\Gamma_1$ and $ \Gamma_2$ that the respective state-space metric
constraints satisfy
\begin{eqnarray}
g_{ii}(X_a)&>& 0 \ \forall \ i \in \{1,2,3,4\} 
\end{eqnarray}

Furthermore, it is intriguing to note that the double center black hole configurations arise
with two different charge vectors have the same set of principle minors. In particular,
we find that both the first center carrying charges, $p0:=1$; $p:=3 \Lambda$; $q:=6 \Lambda^2$;
$q0:=-6 \Lambda$; and the second center carrying charges, $p0:=-1$; $p:=3 \Lambda$;
$q:=-6 \Lambda^2$; $q0:=-6 \Lambda$ have the same principle minors
\begin{eqnarray}
p_1&=& 108 \pi \vert \Lambda \vert^3 (\frac{6 \Lambda^2+ 12 \Lambda^4+ 8 \Lambda^6+ 1}
{(3 \Lambda^4- 1)^{3/ 2}})  \nonumber \\
p_2&=& -972 \pi^2 \vert \Lambda \vert^4 (\frac{1+ 8 \Lambda^2+ 24 \Lambda^4+ 32 \Lambda^6+ 16\Lambda^8}
{(3 \Lambda^4- 1)^{-2}})  \nonumber \\
p_3&=& 972 \pi^3 \vert \Lambda \vert^3 (\frac{6 \Lambda^2+ 12 \Lambda^4+ 8 \Lambda^6+ 1}
{(3 \Lambda^4- 1)^{-3/ 2}})
\end{eqnarray}

Finally, it interesting to note that the general expression of determinant of the metric tensor
implies well-defined state-space manifold $(M_4,g)$, and in addition, we find that the state-space
scalar curvature signifying global correlation properties of underlying statistical system have
no divergence for all non-zero entropy solution. As expected, this confirm for all admissible
values of brane charges that the correlation volume of both the single center solution and double
center solutions modulate as inverse function of the entropy associate with chosen basin of the
$D_6$-$D_4$-$D_2$-$D_0$ multi-centered black brane configurations.

\section{Fractionation of Branes: $D_0$-$D_4$ Black Holes}

The present section as an elucidation of the state-space geometry of small black holes in string
theory carries set of electric charges and a magnetic charge. We notice that the state-space geometry
of these solutions are natural to analyze in the type-$II$ string theory description. In particular,
it is known that these black holes in general may carry finite number of clusters parameters, {\it viz.},
electric charges and magnetic charge which characterize the vacuum string theory configurations
made out of $ D_0 $ branes and $ D_4 $ branes \cite{9707203v1, 0507014v1, 0502157v4, 0505122v2}.
Furthermore, the general details of \cite{0409148v2,AtishHarveyPRL,9511053v1,0511120v2,0410076v2}
have been noteworthy towards some of our subsequent considerations.

In order to make contact of state-space geometry with definite microscopic perspective,
let us consider the chiral primaries of $ SU(1,1 \mid 2)_Z $ and then the associated
supersymmetric ground states of $ \mathcal N= 4 $ supersymmetric quantum mechanics
\cite{0412322} furnishes an understanding of the microscopics of small black holes
and concerned electric brane fractionations. In this consideration, we may easily
see that the are $ 24 p $ bosonic chiral primaries with total $ D_0 $ brane charge
$ N $ in the background with fixed magnetic $ D_4 $ charge $ p $. Then, the degeneracy
involved with the counting of microstates arises from the combinatorics of total $ N $
number of the $ D_0 $ brane charge splitting into $ k $-small clusters with $ n_i $
number of $ D_0 $ branes on each cluster such that the sum $ \sum_{i=1}^k n_i= N $
corresponds to the wrapped $ D_2 $ branes residing on any of the $ 24 p $ bosonic
chiral primary states. Here, the counting is done with the degeneracy $ d_N $ of
states having level function $ N $ in the $ (1+1) $ CFT with $ 24 p $ bosons,
and thus one renders with the celebrated leading order microscopic entropy formula
\begin{eqnarray} \label{fracentropy}
S= \ln d_N= 4 \pi \sqrt{\sum_{i=1}^k n_i p}
\end{eqnarray}

Below, we shall sequentially compute that the components of state-space covariant
metric tensor being defined as the negative Hessian matrix with respect to given
$k$-electric charges $\{ n_i \}_{i=1}^{k} $ on $D_0$-branes and the magnetic charge
$p$ on $D_4$-branes, and thereby divulge the state-space notion of metric positivity,
relative correlations as well as planer and hyper-planer stability for the finite
cluster small black brane configurations. Note that the properties of single cluster
configurations are already considered in the very beginning of the present investigation.

An illustration of the basic idea of state-space geometry of these particular black holes
remains the same as that of the excited string carrying $ n_1 $ number of winding modes and
$ n_p $ number of momentum modes. As we have first considered the case of simplest two charge
extremal configurations, it turns out the same that the state-space geometry of single cluster
configurations can be analyzed in terms of the net electric charges replacing the winding modes
and net magnetic charge replacing the momentum modes of an excited string. In the next subsection,
we shall first consider the two cluster configuration and analyze respective state-space scaling
relations and stability properties.

\subsection{Two Electric Charge Fractionation}
In order to find general pattern of state-space geometric objects of brane fractionated small
black holes, we shall in this subsection explain the state-space geometry for some potential
values by restricting the number of electric clusters in which the total $N$ number of $D_0$
brane charge splits into specific finite partitions. In particular, we shall first explore
the case for k=2 for which the entropy with two clusters takes the form of
\begin{eqnarray}
S= 4 \pi \sqrt{p(n_1+ n_2)}
\end{eqnarray}

In this case, there are two set of charges carried by the small
black holes which can form coordinate charts on underlying
state-space configuration. We shall take he first set of
state-space variables to be the fractionated $D_0$ brane numbers
$\{n_1, n_2 \}$ which are simply proportional to available
fraction of the electric charges present in respective clusters,
while the other state-space variable is the number of $D_4$ brane
which is represented by the magnetic charge $p$. We thence find
that the components of state-space metric tensor arising from the
Hessian of entropy of $D_0$-$D_4$ black holes are
\begin{eqnarray}
g_{p p}&=&  \frac{\pi}{p} \sqrt{\frac{n_1+ n_2}{p}} \nonumber \\
g_{p n_1}&=&
g_{p n_2}=  -\frac{\pi}{\sqrt{p(n_1+ n_2)}} \nonumber \\
g_{n_1 n_1}&=& 
g_{n_1 n_2}= 
g_{n_2 n_2}=  \frac{\pi}{(n_1+ n_2)} \sqrt{\frac{p}{n_1+ n_2}}
\end{eqnarray}

For $i, j \in \{n_1,n_2\}$ and $p$, we observes that the statistical pair correlations
just accounted may in turn be simply ascertained by microscopic descriptions which are
being expressed in terms of large integers (or associated brane charges) of the $D_0$-$D_4$
small black brane solutions connoting an ensemble of microstates. Furthermore, it is evident
for the small black brane configurations that the principle components of the statistical pair
correlations are positive definite for allowed values of the concerned parameters of small black
brane solution. As a result, we can easily see for all admissible set of $n_1$, $n_2$ and $p$ that
the components of state-space metric tensor as given above comply
\begin{eqnarray}
g_{pp} (n_1,n_2,p)> 0, \ 
g_{n_in_i} (n_1,n_2,p)&>& 0 \ \forall \ i= 1,2
\end{eqnarray}

The principle components of state-space metric tensor $\lbrace g_{n_i n_i}, g_{pp} \vert \
i=1,2 \rbrace$ in effect signify a set of positive definite heat capacities (or the related
compressibilities) of the two cluster configurations. In fact, the positivity constraint apprises
that the $D_0$-$D_4$ black branes comply an underlying locally equilibrium statistical configuration.
Furthermore, it is intriguing to note that the non diagonal component $g_{n_in_j}$ also takes a positive
value, \textit{viz.}, we have
\begin{eqnarray}
g_{n_1n_2}&>& 0 \ \forall \ (n_1,n_2,p)
\end{eqnarray}

This shows that the correlations between the associated number of $D_0$ branes in $(1+1)$-CFT
(or the charges in dual description) remains positive in the limit of large electric charges.
This is clearly perceptible because of the fact that the leading order fractionated small black
brane configuration becomes unphysical for these values of the brane parameters.

Interestingly, it follows that the ratios of the principle components of statistical pair
correlations involving electric charges or one electric charge in either correlations are identical,
involving the both electric and magnetic charges vary as inverse square of the connoted charges;
while that of the ratios involving the off-diagonal pair correlations modulate only inversely.
From the above expressions, is is not difficult to visualize for the distinct $i,j \in \lbrace n_1,
n_2 \rbrace $ and the magnetic charge $p$ that the admissible statistical pair correlations
as described above obey the following scaling properties
\begin{eqnarray}
\frac{g_{ii}}{g_{jj}}&=&
\frac{g_{ii}}{g_{ij}}=
\frac{g_{ip}}{g_{jp}}= 1 \nonumber \\
\frac{g_{ii}}{g_{ip}}&=&
\frac{g_{ii}}{g_{jp}}=
\frac{g_{ip}}{g_{pp}}=
\frac{g_{ij}}{g_{ip}}= -(\frac{p}{n_1+ n_2}) \nonumber \\
\frac{g_{ii}}{g_{pp}}&=&
\frac{g_{ij}}{g_{pp}}= (\frac{p}{n_1+ n_2})^2
\end{eqnarray}

Apart from the positivity of principle components of state-space metric tensor,
one in order to accomplish the locally stable statistical configuration demands
that all associated principle minors of the configuration should be positive definite.
It is further not difficult to compute list of the principle minors, \textit{viz.},
$\{p_1,p_2\}$ from the Hessian matrix of associated entropy of fractional $D_0$-$D_4$
black branes. In fact, after some simple manipulations one encounters that the concerned
stability conditions at a point, along one dimensional lines and the two dimensional
surfaces of state-space manifold are respectively measured by
\begin{eqnarray}
p_0= 1, \  
p_1= \frac{\pi}{p} \sqrt{\frac{n_1+ n_2}{p}}, \ 
p_2= 0
\end{eqnarray}

For all physically allowed values of invariant electric-magnetic charges of the $D_0$-$D_4$
black holes, one thus stipulates that the minor constraint $p_1>0$ obliges that the domain
of ascribed magnetic charge must respectively take positive values, while the surface constraint
$p_2=0$ implies that there are no electric-magnetic charges such that the $p_2$ remains positive
real number, and thus the two dimensional subconfigurations of leading order $D_0$-$D_4$ black holes
are not stable. In effect, we can further inspect complete nature of state-space configuration for
the $D_0$-$D_4$ black branes with electric fractionations that the entire stabilities of the system
do not holds for any value of electric and magnetic charges. This follows from the non existence of
positive definite value of the determinant of the metric tensor. In particular, we see easily for all
$i, j \in \{n_1,n_2\}$ and $p$ that the determinant of the state-space metric tensor finds vanishing value.
We thence deduce that the stability of leading order $D_0$-$D_4$ system in two cluster fractionations
is not very ensured over the Gaussian statistical fluctuations.

\subsection{Three Electric Charge Fractionation}

We focus our attention on an extension of state-space analysis for larger number of electric cluster
for the $D_0$-$D_4$ black brane configurations. The exploration begins by considering three clusters
of $D_0$-branes, and single cluster of $D_4$ magnetic brane for the spherical horizon four dimensional
small black hole solutions. What follows here that the magnetic charge is quantified in terms of the number
$D_4$ branes, while that of the electric charges render as the number of brane present in the chosen cluster
of configurations. More precisely, the underlying electric and magnetic charges take large integer values
in terms of the net number of constituent $D_0$ and $D_4$ branes. In turn, one arrives at the simple
quantization condition that the existing charges may be inscribed as fractionated brane configuration.
Such space-time solutions appear quite naturally in the string theory, see for example \cite{0409148v2, AtishHarveyPRL,9511053v1,0511120v2,0410076v2}. In this case, one finds from the general entropy expression
that the three cluster entropy of $D_0$-$D_4$ black branes is given to be
\begin{eqnarray}
S= 4 \pi \sqrt{p(n_1+ n_2+ n_3)}
\end{eqnarray}

Thus, the intrinsic Riemannian geometry as the equilibrium state-space configuration may immediately
be introduced as earlier from the negative Hessian matrix of the entropy of three electric charges and
one magnetic charge extremal small black holes with $D_0$ brane fractionations. We find that the components
of the state-space metric tensor is easily obtained with respect to the underlying electric charges
$ \{n_1,n_2,n_3\}$ and the magnetic charge $p$ as
\begin{eqnarray}
g_{p p}&=& \frac{\pi}{p} \sqrt{\frac{n_1+ n_2+n_3}{p}} \nonumber \\
g_{p n_1}&=&
g_{p n_2}=
g_{p n_3}= -\frac{\pi}{\sqrt{p(n_1+ n_2+n_3)}} \nonumber \\
g_{n_1 n_1}&=& 
g_{n_1 n_2}=
g_{n_1 n_3}=
g_{n_2 n_2}=
g_{n_2 n_3}=
g_{n_3 n_3}= \frac{\pi}{(n_1+ n_2+ n_3)} \sqrt{\frac{p}{n_1+ n_2+ n_3}}
\end{eqnarray}

For all $i, j \in \{n_1,n_2,n_3\}$ and $p$, we notice that the similar set of positivity
conditions and state-space scaling relations are followed as that of the two electric charge
fractionation. Hitherto, we see apparently that the principle components of state-space pair
correlations remain positive definite quantities for all admissible values of underlying electric
magnetic charges of the black brane configuration. It is eassy to observe for given $n_1$, $n_2$,
$n_3$ and $p$ that the following state-space metric constraints are satisfied
\begin{eqnarray}
g_{pp}(n_1,n_2,n_3,p)> 0,\ 
g_{n_in_i}(n_1,n_2,n_3,p)> 0 \ \forall \ i= 1,2,3
\end{eqnarray}

Physically, one may thus note that the principle components of state-space metric tensor
$ \lbrace g_{ii}, g_{pp} \ \vert \ i=n_1,n_2,n_3 \rbrace$ signify a set of heat capacities
(or the associated compressibilities) whose positivity exhibits that the underlying $D_0$-$D_4$
small black brane system is in the locally equilibrium statistical configurations.
Our analysis further complies that the positivity of $g_{pp}$ obliges that the associated
dual conformal field theory living on the boundary must prevail a non-vanishing value of
the magnetic charge defining an associated degeneracy of large number of conformal field
theory microstates. It is worth to mention for given $i, j \in \{n_1,n_2,n_3\}$ and $p$
that the inter cluster state-space correlation functions are again non-trivial in nature.
In particular, we see in this case that the non diagonal components $g_{n_in_j}$ of the
metric tensor take definite positive values
\begin{eqnarray}
g_{n_in_j}(n_1,n_2,n_3,p)&>& 0 \ \forall \ i \neq j \in \{1,2,3\}  
\end{eqnarray}

We may notice further that the ratio of principle components of
state-space pair correlations form three different sets of
relations and specifically we find in a chosen cluster that they
remain the same, vary as inverse of the involved electric magnetic
charges, and vary as inverse square of the involved parameters. It
is in fact not difficult to inspect for non-identical $i,j,k \in
\lbrace 1,2,3 \rbrace $ and $p$ that the state-space pair
correlations are consisting of the following type of scaling
relations
\begin{eqnarray} \label{frac3e}
\frac{g_{ij}}{g_{jj}}&=&
\frac{g_{ii}}{g_{jj}}=
\frac{g_{ik}}{g_{ii}}=
\frac{g_{ik}}{g_{jj}}=
\frac{g_{ik}}{g_{ij}}=
\frac{g_{ip}}{g_{jp}}= 1 \nonumber \\ 
\frac{g_{ii}}{g_{ip}}&=&
\frac{g_{ij}}{g_{jp}}=
\frac{g_{ij}}{g_{kp}}=
\frac{g_{ii}}{g_{jp}}=
\frac{g_{ip}}{g_{pp}}= -(\frac{p}{n_1+ n_2+n_3}) \nonumber \\
\frac{g_{ii}}{g_{pp}}&=&
\frac{g_{ij}}{g_{pp}}= (\frac{p}{n_1+ n_2+n_3})^2 
\end{eqnarray}

An investigation of definite global properties of three electric clustered $D_0$-$D_4$ black brane
configurations determines certain stability consideration along each directions, each planes
and each hyper-planes, as well as on the entire intrinsic state-space manifold. Specifically,
we can determine whether the underlying $D_0$-$D_4$ configuration is locally stable on state-space
planes and hyper-planes, and thus one need to compute corresponding principle minors of negative
Hessian matrix of the $D_0$-$D_4$ black hole entropy. In this case, we may easily appraise for
all physically likely values of magnetic charge and electric charges that the possible principle
minors computed from the above state-space metric tensor are
\begin{eqnarray}
p_0= 1, \  
p_1= \frac{\pi}{p} \sqrt{\frac{n_1+ n_2+n_3}{p}}, \ 
p_i= 0, \ i= 2,3 
\end{eqnarray}

In the entropy representation, it could thus be seen that the principle minors defined by
\begin{eqnarray}
p_2(n_1,n_2,n_3,p)&:=& g_{11} g_{22}-g_{12}^2   \nonumber  \\
p_3(n_1,n_2,n_3,p)&:=& g_{n1n1}(g_{n2n2} g_{n3n3}-g_{n2n3}^2)- g_{n1n2} (g_{n1n2}g_{n3n3}  \nonumber  \\&&
- g_{n1n3} g_{n2n3})+ g_{n1n3} (g_{n1n2} g_{n2n3} -g_{n1n3} g_{n2n2})
\end{eqnarray}

vanish identically for all admissible values of the electric charges $n_1$,$n_2$, $n_3$
and magnetic charge $p$. In turn, one can easily observe that the vanishing condition
$p_{i>1}(n_1,n_2,n_3,p)= 0$ signifying the state-space configurations corresponding to the
three clusters of electric $D_0$-branes indicate that the statistical system remains
unstable over possible surfaces and hyper-surfaces. Furthermore, we indeed find for
entire system that the positivity of final minor is just the positivity condition of the
determinant of metric tensor. Thence, an easy inspection observes further that the
determinant of the metric tensor vanishes as well for all three clusters of electric charges
and magnetic charge $\{n_1,n_2,n_3,p\}$ which form co-ordinates on its state-space configuration.

\subsection{Multi Electric Charge Fractionation}

Now we shall consider the state-space configuration for the most general case of brane
fractionation in the finite cluster of $D_0$-$D_4$ small black branes, and present our analysis
from the view-points of associated microscopic entropy obtained for $k$-clusters. It turns out
that the involved entropy can be defined via an appropriate degeneracy formula, and the concerned
expression reduces to the entropy as ascribed in Eqn. (\ref{fracentropy}).

The state-space geometry describing the local pair correlations between the equilibrium microstates
of multi-clustered charged extremal $D_0$-$D_4$ black holes resulting from degeneracy of microstates
may thence be computed as earlier from the Hessian matrix of Eqn. (\ref{fracentropy}) with respect to the
parameters, \textit{viz.}, the $D_0$ electric charges $\{n_1,n_2,\ldots n_k\}$ and the $D_4$ magnetic
charge $p$. At this juncture, we obtain that the components of underlying state-space covariant metric
tensor are generically given by
\begin{eqnarray}
g_{p p}&=& \frac{\pi}{p} \sqrt{\frac{\sum_{i=1}^k n_i}{p}} \nonumber \\
g_{p n_i}&=& -\frac{\pi}{\sqrt{p(\sum_{i=1}^k n_i)}} \nonumber \\
g_{n_i n_j}&=& \frac{\pi}{(\sum_{i=1}^k n_i)} \sqrt{\frac{p}{\sum_{i=1}^k n_i}},\ \forall i,j= 1, 2,\ldots, k
\end{eqnarray}

For finite number of parameters of the $D_0$-$D_4$ black brane configurations, \textit{viz.},
the electric charges $i, j, k, l \in \{n_1,n_2,\ldots n_k\}$ and the magnetic charge $p$,
we observe that the specific inspections observed in previous subsections for the two and
three cluster systems hold, as well. In effect, it is evident in general that the principle
components of equilibrium statistical pair correlations are positive definite for all allowed
values of concerned parameters of the $D_0$-$D_4$ small black branes in each electric clusters.
As an immediate result, one finds from the present analysis that the concerned state-space
metric constraints are satisfied with
\begin{eqnarray}
g_{n_in_i}&>& 0 \ \forall \ i= 1,2,\ldots,k \nonumber \\
g_{pp}&>& 0 \ \forall \ (n_1,n_2,\ldots,n_k,p)
\end{eqnarray}

Interestingly, it is worth to mention that our geometric expressions arising from the entropy of small
black holes indicate that some of the brane charges are possible to be safely turn off, say $n_i= 0$,
while having a well-defined state-space geometry. However, it is unfeasible to have an intrinsic
state-space configuration of small black holes with no electric charge or no magnetic charge,
say $n_i= 0 \forall i$ or $p= 0$, since the objects inside the square-root of the statistical
entropy vanishes, and thus the argued small black hole configurations with vanishing number of
either $D_0$ branes or $D_4$ branes are no more well-defined state-state configuration. The case
of finitely many electric branes indeed agrees with our expectation that the non diagonal components
$g_{n_in_j}$ find their respective positive values
\begin{eqnarray}
g_{n_in_j}&>& 0 \ \forall \ (n_1,n_2,\ldots,n_k,p)
\end{eqnarray}

Under the present considerations, we thus observe for a given
fraction of the electric charges $ i \neq j \neq k \neq l \in
\lbrace n_1,n_2,\ldots,n_k \rbrace $ and the magnetic charge $p$
that the relative state-space pair correlation functions form the
same scaling qualifications as in the case of three clusters of
$D_0$ electric branes delt in Eqn. (\ref{frac3e}), except the fact
that now the sum in the denominator runs over $\{1,2,\cdots,k \}$.
Furthermore, one may now easily see for the $D_0$-$D_4$
configurations involving four or higher clusters of $D_0$-branes
that there exist an extra identical scaling relation
\begin{eqnarray}
\frac{g_{ij}}{g_{kl}}=1
\end{eqnarray}

We thus see for the most general leading order brane fractionation in $D_0$-$D_4$ system
that there are total 14 type of relative correlation functions at chosen state-space basis.
It is worth to mention that an appraisal of exhaustive state-space stability constraints
demands that all the associated principle minors must be positive definite, as the positivity
of principle components of metric tensor defines the local linear stability in the neighborhood
of chosen local co-ordinate chart on an underlying $(M_{k+1},g)$ describing concerned state-space
manifold of finite clustered $D_0$-$D_4$ solutions.

Specifically for $i, j \in \{n_1,n_2,\ldots n_k\}$ and $p$, we find that the are no extra
type of planer and hyper-planer stabilities as that of the relative state-space correlation
functions than the linearly stable multi-clustered $D_0$-$D_4$ small black brane system.
It is rather easy to divulge the physical picture of the solution set and in fact after
some simplifications one discovers that the planer stability criteria on the two dimensional
surfaces and hyper-planer stability criteria on the three or higher dimensional surfaces of
the state-space manifold may simply be rendered from the definition of the state-space geometry.

Intriguingly, it is not difficult to compute from the consideration of the Hessian matrix
of $k$-clustered $D_0$-$D_4$ black brane leading order entropy solutions that the list of
non-zero principle minors remains the same as that of the two or three clustered configurations.
In addition, as in the case of two and three clusteres of electric charges, we observe for
general $k$ electric charge configurations that the set of all possible principle minors
$\{ p_i(n_1,n_2,\ldots n_k,p) \vert \ \forall i> 1 \}$ remain zero on the state-space manifold
$(M_{k+1},g)$ as well as on respective lower dimensional associated systems of multi-clustered
$D_0$-$D_4$ black branes.

It is worth to mention in particular that the local stability of full small black brane
state-space configuration is determined by computing the determinant of concerned
state-space metric tensor. Herewith, we may in principle as well compute compact formula
for the determinant of the metric tensor, and indispensably, our intrinsic state-space
geometric analysis arising from the leading order entropy consideration demonstrates
that the the determinant of the metric tensor do not find a non-vanishing value for
any admissible finite electric clusters of the $D_0$-$D_4$ black brane configurations.

In the next section, we shall consider implications of state-space geometry arising
from the Fuzzball solutions, and explicate the nature of scaling properties of possible
state-space pair correlation functions and stability requirements of the fuzzy ring solutions
in the setup of Mathur's fuzzball consideration \cite{0706.3884v1}.

\section{The Fuzzball Solutions: Fuzzy Rings}

The view-points of the Mathur's fuzzball solutions \cite{0502050v1} are considered
in this section. To be specific, we shall analyze concerned aspects of state-space
geometry for the most exhaustively studied two charge extremal black branes having
electric-magnetic charges, $(Q,P)$ and an angular momentum $J$. We shall focus in
particular to analyze the state-space observations in terms of concerned parameters
of the fuzzball solution, and thereby shed light on the state-space quantities from
Mathur recent proposal to find ensemble of microstates which form an equilibrium
statistical basic over which we shall define associated thermodynamic intrinsic
state-space geometry.

It is worth to mention in the fuzzball picture that one can construct classical space-time
geometry with definite horizon topology when many of the quanta of underlying three
parameter $D_1$-$D_5$-$P$ CFT lie in the same mode. Nevertheless, it turns out in
general that the generic states will not have all the quanta placed in a few modes,
so the throat of concerned black hole space-times ends in a very quantum fuzzball,
see for introduction of the fuzzball solutions \cite{0706.3884v1,0109154v1,0202072v2}.

It is however interesting to note in the fuzzball picture that the actual
microstates of such black branes do not have an event horizon, but it is
rather the area of the boundary of fuzzy region where microstates start differing
from each other satisfies Bekenstein-Hawking type relation and thereby defines
an entropy inside the chosen boundary. Moreover, in turns out according to
string theory picture that the different microstates are `cap off' before
reaching the end of an infinite throat and thus they give rise to different
near horizon space-time geometries. In particular, the average throat behaves
as inverse of average radius of the fuzzballs. Thus the Bekenstein-Hawking entropy
\cite{0706.3884v1,0109154v1,0202072v2} has been obtained from the area of such
a stretched horizon whose state-space interpretation may be obtained from the
coarse graining statistical entropy
\begin{eqnarray}
S(Q,P,J)= C \sqrt{QP- J}
\end{eqnarray}

The associated state-space geometry of rotating two charge fuzzy ring system can
thence be constructed out of the parameters which characterize the microstates
of black brane. In particular, we can perform an investigation either in terms
of the $ D_1 $ brane electric charge $ Q $ and $ D_5 $ brane magnetic having charge
$ P $ or correspondingly $ n_1 $ number of $ D_1 $ branes and $ n_5 $ number of
$ D_5 $ branes. Then, the dimension of state-space manifold is equal to the number
of actual parameters which defines the fuzzy black ring solution. We shall thence
study the state-space configurations whose co-ordinates deal with the charges or
number of constituent branes. In particular, we shall consider the electric-magnetic
charges $(Q,P)$ and angular momentum $J$ that they define co-ordinates on concerned
state-space manifold of the two charge fuzzy black ring solution.

The state-apace geometry constructed out of equilibrium state of the rotating two charged
extremal black ring resulting from the entropy can now easily be computed as earlier from
the negative Hessian matrix of the entropy with respect to the charges and angular momentum.
Note that an understanding of the state-apace geometry based on the stretched horizon requires
the classical time scale limit of fuzzball. This is because the size of fuzzball is made by
the generic states such that its surface area in leading order satisfies Bekenstein-Hawking
type relation with the entropy of the fuzzball whose boundary surface becomes like a horizon
only over classical time scales. We may thefore see that the components of the metric tensor
are explicitly given as
\begin{eqnarray}
g_{PP}&=& \frac{1}{4}CQ^2(PQ-J)^{-3/2},\ 
g_{PQ}= -\frac{1}{4}C(PQ- 2J)(PQ-J)^{-3/2} \nonumber \\
g_{PJ}&=& -\frac{1}{4}CQ(PQ-J)^{-3/2}, \ 
g_{QQ}= \frac{1}{4}CP^2(PQ-J)^{-3/2} \nonumber \\
g_{QJ}&=& -\frac{1}{4}CP(PQ-J)^{-3/2}, \ 
g_{JJ}= \frac{1}{4}C(PQ-J)^{-3/2}
\end{eqnarray}

From the simple $D$-brane description, we observe that there exists an interesting
brane interpretation which describe the state-space correlation formulae arising
from the corresponding microscopic entropy of the aforementioned two charge rotating
$D_1$-$D_5$ solutions. Furthermore, the state-space correlations turn out to be in
precise accordance with an associated attractor configuration being disclosed in the
limiting special Bekenstein-Hawking solution. In the entropy representation, it has
thence been noticed that the Hessian matrix of the entropy illustrates basic nature
of possible state-space correlations between the set of extensive variables which in
this case are nothing more than the $D_1$ and $D_5$-brane charges and angular momentum.
As mentioned before, we can articulate in this case as well that for all non-zero
admissible values of $P,Q,J$, the principle components of intrinsic state-space
metric tensor satisfy
\begin{eqnarray}
g_{PP}> 0,\ 
g_{QQ}> 0,\ 
g_{JJ}> 0
\end{eqnarray}

Substantially, the principle components of state-space metric tensor signifies
heat capacities or the associated compressibility whose positivity indicates that
the underlying statistical system is in local equilibrium consisting of the $D_1$
and $D_5$-brane configurations. Furthermore, we perceive that the ratio of possible
diagonal components varies as inverse square, which weaken faster and thus relatively
quickly come in to an equilibrium configuration, than those involving the off diagonal
components varying inversely in the involved parameters. Incidentally, the ratios of
non-diagonal components varying inversely remain comparable for a longer domain of
parameters varying under the Gaussian fluctuations. We have in particular inspected
$ \forall i \neq j \in \lbrace P,Q \rbrace $ and $J$ that the relative pair correlation
functions satisfy the following scaling relations
\begin{eqnarray}
\frac{g_{ii}}{g_{jj}}&=& (\frac{j}{i})^2,\ 
\frac{g_{ii}}{g_{JJ}}= j^2,\ 
\frac{g_{ij}}{g_{ii}}= -\frac{1}{j^2}(PQ-2J)\nonumber \\
\frac{g_{ii}}{g_{iJ}}&=& -j,\ 
\frac{g_{iJ}}{g_{jJ}}= \frac{j}{i},\ 
\frac{g_{ii}}{g_{jJ}}= - \frac{j^2}{i}\nonumber \\
\frac{g_{iJ}}{g_{JJ}}&=& -j,\ 
\frac{g_{ij}}{g_{iJ}}= \frac{1}{j}(PQ-2J),\ 
\frac{g_{ij}}{g_{JJ}}= - (PQ-2J)
\end{eqnarray}

An investigation of definite global properties of two charged Fuzzball configurations determines
certain stability approximation along each directions, each planes, each hyper-planes and on entire
intrinsic state-space manifold. In this case, as we intend to determine whether the underlying Fuzzball
configuration is locally stable on state-space planes and hyper-planes, and thus we are required
to compute corresponding principle minors of negative Hessian matrix of the entropy. Specifically,
we may easily appraise for all physically likely values of brane charge and angular momentum that
the possible principle minors computed from the above state-space metric tensor are non-zero and
definite function of the electric-magnetic charges $\{P,Q \}$ and an angular momentum $J$.
We in effect see for all admissible parameters describing the three parameter fuzzball solutions
that the list of concerned state-space stability functions is
\begin{eqnarray}
p_1= \frac{1}{4}  C Q^2 (P Q- J )^{-3/ 2},\ 
p_2= \frac{1}{4}  C^2 J (P Q- J )^{-2}
\end{eqnarray}

Thus, the minor constraints on $p_1,p_2$ implies that the two charge Fuzzball solution under
consideration are stable over the lines, planes of the state-space configuration for all values
of the $D_1$, $D_5$ brane charges and any positive value of the angular momentum. As we have shown
in the previous examples that the determinant of the metric tensor thus defined is non-zero for
non-zero brane charges and angular momentum. In fact, it is easy to observe that the determinant
of the metric tensor reduces to
\begin{eqnarray}
\Vert g \Vert= -\frac{1}{16}C^3 (PQ- J)^{-5/2}
\end{eqnarray}

Similarly, the constraint $p_3:=g(Q,P,J)<0$ results in an interpretation that this configuration
is globally unstable over the full intrinsic state-space configurations. This is also intelligible
from the fact that the responsible equilibrium entropy tends to its maximum value, while the same
culmination do not remain valid over the entire state-space manifold. It may in turn be envisaged
in the $D_1$-$D_5$-$P$ description that the fuzzball black rings do not correspond to intrinsically
stable statistical basis when all the configuration parameters fluctuate. Thus, it is very probable
that the underlying ensemble of chosen CFT microstates upon subleading higher derivative corrections
may smoothly move into the more stable brane configurations.

Finally, in order to elucidate universal nature of the statistical interactions and the
other properties concerning fuzzball rotating black rings, one needs to determine definite
global state-space geometric invariant quantities on its intrinsic state-space manifold.
Indeed, we notice that the indicated simplest invariant is achieved just by computing
the state-space scalar curvature, which as explained in \cite{BNTBull} be obtained in
straightforward fashion by applying the standard method of our intrinsic geometry.
In the large charge limit in which the asymptotic expansion of the entropy of the two
charge rotating ring solution is valid, we notice in particular that the state-space
scalar curvature can rather be expressed as an inverse function of the entropy.

An exact analysis in turn finds that the constant of proportionality between the state-space
scalar curvature and conting entropy to be a negative constant, and thus we find fuzzy ring to
be an attractive statistical configuration, see for related interpretations \cite{BNTBull}.
Most importantly, it turns out in the limit when the fuzzy ring is viewed in the perspective of many
fuzzballs that the present analysis relies on corrected averaged horizon configuration. Finally,
it is worth to mention that the statistical systems of the $D_1$-$D_5$-$P$ fuzzy rings find an
intriguing conclusion in the Gaussian approximations, and consequently, the present description
vindicates physically sound containments that the state-space configuration of fuzzy rings is
non-degenerate, curved and an everywhere regular intrinsic Riemannian manifold $(M_3,g)$.

\section{Bubbling Black Brane Solutions: Black Brane Foams}

In this section, we finally analyze the state-space geometry of an ensemble of equilibrium
microstates characterizing three charge foamed black brane configurations in $M$-theory \cite{0604110}.
These supergravity bubbling solutions naturally appears in the string theory and $M$-theory,
see for concerned details \cite{0604110, 0408120v2, 0409174v2, 0505166v2}. The study of bubbled
space-time geometries and axi-symmetric merger solutions thence turn out to be interesting to
investigate further from the view-points of our state-space geometry. We shall here show that
the possible characterization of state-space geometry has herewith been accomplished in terms
of the parameters describing an ensemble of microstates for the three charge black brane foam
solutions \cite{07063786v2}.

\subsection{A Toy Model: Single GH-center}

The state-space geometry arising from entropy of the foam configurations having single
Gibbon-Hawking center can be divulged by considering $M$-theory background \cite{0604110}
compactified on $T^6$. In the large N limit, one thence finds set of flux parameters which
may be written in terms of brane charges. It is worth to remark that the associated topological
entropy which is independent of the number and charges on the Gibbons-Hawking base points.
The origin of such an entropy lies solely in the possible number of the choices of positive
quantized fluxes on topologically non-trivial cycles.

In effect, it turns out that these cycles satisfy definite constraints, \textit{viz} one finds
in particular that the supergravity and worldvolume descriptions have the same relation between
the brane parameters which determined the entropy of the bubbled black brane foam, see for instance
\cite{0604110}. Note that an understanding of the state-apace geometry based on the bubbling black
branes require knowledge of Bekenstein-Hawking entropy which could be obtained from the area of the
horizon of chosen solution. A microscopic interpretation may thence be offered as coarse graining
of concerned combinatorial entropy of the foam.

To describe state-space geometry of single center Gibbons-Hawking configuration, we shall
in particular consider set of flux parameters $ \lbrace k_i^1, k_i^2, k_i^3 \rbrace $ to be
positive half integers \cite{0604110}, and then the topological entropy coming from the leading
order contributions of the fluxes $ \lbrace k_i^1 \rbrace $ where the index $i$ defines positions
of the Gibbons-Hawking base points has been written as
\begin{eqnarray}
S(Q_1, Q_2, Q_3):= \frac{\pi}{3}\sqrt{6}(\frac{Q_2 Q_3}{Q_1})^{1/4}
\end{eqnarray}

As proclaimed in the previous subsections, we notice in this case as well that the state-apace
geometry describing the nature of equilibrium brane microstates can be constructed out of the
three charges of the bubbled black brane foams. The covariant metric tensor as invoked earlier
can immediately be computed from negative Hessian matrix of the foam entropy resulting from underlying
statistical configuration. Thus, the brane charges, \textit{viz.}, $ \lbrace Q_1, Q_2, Q_3 \rbrace $
form the coordinate charts for the state-space manifold of our interest and thus with respect to the
brane parameters we may describe the typical intrinsic geometric features of the bubbled black brane
foams having single GH center. In fact, we notice that the components of the covariant metric tensor
can easily be presented to be
\begin{eqnarray}
g_{Q_1Q_1}&=& -\frac{5 \pi \sqrt{6}}{48 Q_1^2} (\frac{Q_2 Q_3}{Q_1})^{1/4},\ 
g_{Q_1Q_2}=  \frac{\pi \sqrt{6} Q_3}{48 Q_1^2}(\frac{Q_1}{Q_2 Q_3})^{3/4} \nonumber \\
g_{Q_1Q_3}&=&  \frac{\pi \sqrt{6} Q_2}{48 Q_1^2}(\frac{Q_1}{Q_2 Q_3})^{3/4},\ 
g_{Q_2Q_2}=  \frac{\pi \sqrt{6}}{16}(\frac{Q_3}{Q_1})^{2} (\frac{Q_1}{Q_2 Q_3})^{7/4} \nonumber \\
g_{Q_2Q_3}&=& - \frac{\pi \sqrt{6}}{48 Q_1}(\frac{Q_1}{Q_2 Q_3})^{3/4},\ 
g_{Q_3Q_3}=  \frac{\pi \sqrt{6}}{16}(\frac{Q_2}{Q_1})^{2} (\frac{Q_1}{Q_2 Q_3})^{7/4}
\end{eqnarray}

One thus appreciates for all $i, j,k \in \{1,2,3\}$ describing the single GH center
bubbling brane configuration that the state-space geometry materializing from the leading
order Bekenstein-Hawking entropy of the toroidally compactified $M$-theory configuration
admits remarkably simple expressions in terms of physical charges. It may again be expected
that the microscopic preliminaries would plausibly be suggested via the Cardy formula or
associated general Hardy-Ramanujan formula. As enumerated in the earlier sections, we
nevertheless stress out for all non-zero values of the brane charges $Q_1,Q_2,Q_3 $ that
the principle components of concerned state-space metric tensor satisfy
\begin{eqnarray}
g_{Q_1Q_1} < 0,\ 
g_{Q_2Q_2} > 0,\ 
g_{Q_3Q_3} > 0
\end{eqnarray}

The present analysis physically proclaims that the principle components of state-space metric
tensor signify heat capacities or the relevant compressibilities whose positivity connote that
the underlying statistical system is in locally stable equilibrium configurations of an ensemble
of dual CFT microstates. Moreover, it is rather instructive to note that the behavior of brane-brane
statistical pair correlation defined as $g_{Q_1Q_1}$ is asymmetric in contrast to the other existing
diagonal correlations. In fact, one can understand it by arguing that an increment of the $Q_1$ brane
charge reduces the entropy and thus it correspond to locally unstable state-space interactions than
the brane-brane self-interactions involving either $Q_2$ or $Q_3$ charges.

Furthermore, it has been substantiated that the ratio of diagonal components varies
as inverse square of the invariant parameters which vary under the Gaussian fluctuations,
whereas the ratios involving off diagonal components vary only inversely in the chosen charges.
In particular, we see for $i, j,k \in \{Q_1,Q_2,Q_3\}$ describing the single GH center configuration
that the possible relative state-space correlation functions are defined as
\begin{eqnarray}
C_{BB}:= \{ \frac{g_{ij}}{g_{jj}}, \frac{g_{ii}}{g_{jj}}, \frac{g_{ik}}{g_{ii}}, \frac{g_{ik}}{g_{jj}}, \frac{g_{ik}}{g_{ij}} \}
\end{eqnarray}

This suggests that the diagonal components weaken faster and relatively quickly come into an
equilibrium, than the off diagonal components which remain comparable for the longer domain
of the parameters defining the single GH center bubbling configurations. An explicit observation
shows that the relative pair correlation functions satisfy a simple set of scaling relations.
In particular, we can easily observe for given distinct $i, j, k \in \lbrace Q_1,Q_2,Q_3 \rbrace $
that the possible relative state-space correlation functions for the single GH-center find the
following values
\begin{eqnarray}
C_{BB}^S= \{ \frac{g_{13}}{g_{22}},\frac{g_{12}}{g_{13}},\frac{g_{12}}{g_{22}},\frac{g_{22}}{g_{33}},
\frac{g_{23}}{g_{22}} \}= \{ \frac{1}{3} \frac{Q_2^2}{Q_1Q_3}, \frac{Q_3}{Q_2}, \frac{1}{3} \frac{Q_2}{Q_1},(\frac{Q_3}{Q_2})^2,-\frac{1}{3} \frac{Q_2}{Q_3}\}
\end{eqnarray}

As noticed in the previous configurations, it is not difficult to analyze the local stability
for the bubbling black holes, as well. In particular, one can determine the principle minors
associated with the state-space metric tensor and thereby demand that all the principle minors
must be positive definite. In this case, we may adroitly compute the principle minors from the
Hessian matrix of associated entropy concerning the three charge bubbling black holes.
In fact, after simple manipulation, we discover that the local stability criteria on
the two dimensional surfaces and the three dimensional hyper-surfaces of the underlying
state-space manifold are respectively given by the following relations
\begin{eqnarray}
p_1= -\frac{5\sqrt{6}\pi}{48} Q1^{-9/4} Q2^{1/4} Q3^{1/4},\ 
p_2= -\frac{\pi^2}{24} Q1^{-5/2} Q2^{-3/2} Q3^{1/2}
\end{eqnarray}

For all physically admitted values of brane charges of the bubbling black holes, we may thus
easily ascertain that the minor constraint, {\it viz.}, $p_1(Q_i)>0$ inhibits the domain of
assigned charges that the two of them must be positive and third be negative real number,
while the constraint $p_2(Q_i)>0$ imposes that the brane charges must respectively satisfy above
definite brane charge conditions. In particular, these constraints enables us to investigate the
nature of the state-space geometry of $M$-theory bubbling black holes. We thus observe that
the presence of planer and hyper planer instabilities exist for the bubbling black holes, which
together demand for a restriction on the allowed value of the brane charges.

As stated earlier, we find in this case that the determinant of state-space geometry describing
correlations between two chosen microstates of the bubbled black brane foams may be characterized
in terms of the extensive brane charges of the single GH center solution. Employing state-space
consideration of negative Hessian matrix of the foam entropy with respect to the brane charges
$ \lbrace Q_1,Q_2,Q_3 \rbrace $, we find that the determinant of the metric tensor is given by
\begin{eqnarray}
\Vert g \Vert= -\frac{\pi^3 \sqrt{6}}{384 Q_1^4}(\frac{Q_1}{Q_2 Q_3})^{-5/4}
\end{eqnarray}

Furthermore for equal values of charges $Q_1:=Q$, $Q_2:=Q$ and $Q_3:=Q$; it is easy to see
that the principle minor $p_1:= g_{11}$ reduces to $ p_1= -\frac{5\sqrt{6}\pi}{48} Q^{-7/4} $,
while the surface minor $p_2:= g_{11} g_{22}- g_{12}^2$ shows further that the two dimensional
state-space configurations of underlying single GH center solutions are unstable. In particular,
we find an explicit expression for equal value of charges that the surface minor is given by
$ p_2(Q)= -\frac{\pi^2}{24} Q^{-7/2} $.

As expected, we see for equal value of brane charges, \textit{viz.}, $Q_i=Q$ that the toy model
single GH bubble black brane solution remains unstable over an entire fluctuating statistical
configuration. This follows from the fact that the determinant of the metric tensor as being
the highest principle minor $p_3$ reduces to $g(Q)= -\frac{\pi^3 \sqrt{6}}{384} Q^{-21/4}$.

Interestingly, it is note worthy from the general expression of the determinant of the metric
tensor and in addition that of the state-space scalar curvature signifying global correlation
volume of underlying statistical system that the single GH center bubbled systems are unstable
and finds an attractive statistical nature for given non zero entropy solution. Finally for all
admissible values of brane charges, we come up with the fact that the state-space scalar curvature
signifying global correlation length of an underlying statistical system confirms no divergence,
and in turn it varies as an inverse function of the entropy of chosen single center GH center bubbled
configurations.

\subsection{Black Brane Foams}

In this subsection, we shall consider state-space geometry of the
most general three center GH solutions, which may exhaustively be
contemplated by three brane charges of the bubbled black brane
configurations. The coordinate chart of underlying intrinsic
state-space manifold naturally emerge from the parameters of
equilibrium microstates of chosen bubbling supergravity solutions
\cite{MT,pa-mar,bho,bhko}. It turns out from the details of brane
parameters that one may easily ascribe the state-space definitions
to the central charge contributions associated with the rotating
black branes in Minkowski space, as well. In effect, our attention
shall therefore be focused on possible U-dual configurations, and
describe promising analysis in the viewpoints of
\cite{lu-ma,dghw,ch-oh,mnt}.

Here, very purpose would thus be to exploit the state-space
meanings of symmetric factors of brane charges arising from an
elementary conformal field theory living on the boundary. As we
have encountered the state-space geometry of the single GH center
bubbled black branes in the previous subsection, in this
subsection we shall analyze the state-space fluctuations for
unrestricted $3$-charge bubbled black brane foam solutions
\cite{0604110}. Thereby, we shall examine general nature of
concerned state-space configurations over leading order symmetric
charge contributions into topological entropy of the three charged
bubbled black brane foams characterized by the charges $Q_1$,
$Q_2$ and $Q_3$. It turns out by considering appropriate factors
coming from the partitioning of concerned flux parameters,
\textit{viz.}, $ \lbrace k_i^1, k_i^2, k_i^3 \rbrace $ that the
involved topological entropy may be defined by the following
formula:
\begin{eqnarray}
S(Q_1, Q_2, Q_3):= \frac{2\pi}{\sqrt{6}}\lbrace (\frac{Q_2 Q_3}{Q_1})^{1/4}
+ (\frac{Q_1 Q_2}{Q_3})^{1/4}+ (\frac{Q_1 Q_3}{Q_2})^{1/4} \rbrace
\end{eqnarray}

It is again not difficult to explore the state-space geometry of the equilibrium microstates
of the three charge bubbled black brane foams arising from the entropy expression which concerns
just the Einstein-Hilbert action. As stated earlier that the Ruppeiner metric on the state-space
manifold is given by the negative Hessian matrix of the ring entropy with respect to the thermodynamic
variables. The state-space variables in this case are the conserved brane charges which in turn are
proportional to the fluxes carried by the constituent branes. Explicitly, we find in this case that
the components of covariant state-space metric tensor are
\begin{eqnarray}
g_{Q_1Q_1}&=& -\pi \lbrace \frac{5 \sqrt{6}}{48 Q_1^2}(\frac{Q_2 Q_3}{Q_1})^{1/4}
- \frac{ \sqrt{6}}{16 Q_1}(\frac{Q_2}{Q_3 Q_1})^{1/4}
- \frac{ \sqrt{6}}{16 Q_1}(\frac{Q_3}{Q_2 Q_1})^{1/4}\rbrace \nonumber \\
g_{Q_1Q_2}&=& -\pi \lbrace - \frac{ \sqrt{6}Q_3}{48 Q_1^2}(\frac{Q_1}{Q_2 Q_3})^{3/4}
+ \frac{ \sqrt{6}}{48 Q_3}(\frac{Q_3}{Q_1 Q_2})^{3/4}
- \frac{ \sqrt{6}Q_3}{48 Q_2^2}(\frac{Q_2}{Q_1 Q_3})^{3/4} \rbrace \nonumber \\
g_{Q_1Q_3}&=& -\pi \lbrace - \frac{ \sqrt{6}Q_2}{48 Q_1^2}(\frac{Q_1}{Q_2 Q_3})^{3/4}
- \frac{ \sqrt{6}Q_2}{48 Q_3^2}(\frac{Q_3}{Q_1 Q_2})^{3/4}
+ \frac{ \sqrt{6}}{48 Q_2}(\frac{Q_2}{Q_1 Q_3})^{3/4} \rbrace \nonumber \\
g_{Q_2Q_2}&=& -\pi \lbrace \frac{5 \sqrt{6}}{48 Q_2^2}(\frac{Q_1 Q_3}{Q_2})^{1/4}
- \frac{ \sqrt{6}}{16 Q_2}(\frac{Q_3}{Q_1 Q_2})^{1/4}
- \frac{ \sqrt{6}}{16 Q_2}(\frac{Q_1}{Q_3 Q_2})^{1/4}\rbrace \nonumber \\
g_{Q_2Q_3}&=& -\pi \lbrace \frac{ \sqrt{6}}{48 Q_1}(\frac{Q_1}{Q_2 Q_3})^{3/4}
- \frac{ \sqrt{6}Q_1}{48 Q_3^2}(\frac{Q_3}{Q_1 Q_2})^{3/4}
- \frac{ \sqrt{6}Q_1}{48 Q_2^2}(\frac{Q_2}{Q_1 Q_3})^{3/4} \rbrace \nonumber \\
g_{Q_3Q_3}&=& -\pi \lbrace \frac{5 \sqrt{6}}{48 Q_3^2}(\frac{Q_1 Q_2}{Q_3})^{1/4}
- \frac{ \sqrt{6}}{16 Q_3}(\frac{Q_2}{Q_1 Q_3})^{1/4}
- \frac{ \sqrt{6}}{16 Q_3}(\frac{Q_1}{Q_2 Q_3})^{1/4}\rbrace
\end{eqnarray}

It follows from the above expressions that the statistical pair correlations thus described
can in turn be accounted by a simple geometric descriptions expressed in terms of the brane charges
connoting an ensemble of fluxes for the general three GH centered bubbling black brane configurations.
Furthermore, we observe that the principle components of underlying state-space configuration are
positive definite for all allowed values of the bubbling parameters of the multi center GH solution.
In particular, it is evident for functions $f_{ii}(Q_1,Q_2,Q_3)$ as defined in Eqn. (\ref{fij}) that the
state-space metric constraints defining positivity of concerned diagonal pair correlation functions are
\begin{eqnarray} \label{multi3GHpositity}
g_{Q_iQ_i}(Q_1,Q_2,Q_3)&>& 0 \ \forall \ i \in \{1,2,3\} \mid f_{ii} <0,\ 
\end{eqnarray}

Essentially, the principle components of state-space metric tensor $\lbrace g_{Q_i Q_i} \vert \
i=1,2,3 \rbrace$ signify a set of definite heat capacities (or the related compressibilities ) whose
positivity for a range of involved charges as presented below apprises that the bubbled black holes
comply an underlying locally stable equilibrium statistical configurations along each directions.
It is intriguing to note that the positivity of $g_{Q_iQ_i}$ holds even if some of the brane charges
of the associated brane charges become zero. This is clearly perceptible because of the fact that the
brane configuration remains physical and locally stable for all brane charges $(Q_1,Q_2,Q_3)$ such that
the following relations defining Eqn. (\ref{multi3GHpositity}) are satisfied
\begin{eqnarray} \label{fij}
f_{11}(Q_1, Q_2, Q_3)&:=& \frac{5 \sqrt{6}}{48 Q_1^2} (\frac{Q_2 Q_3}{Q_1})^{1/4}- \frac{ \sqrt{6}}{16 Q_1}
(\frac{Q_2}{Q_3 Q_1})^{1/4}- \frac{ \sqrt{6}}{16 Q_1}(\frac{Q_3}{Q_2 Q_1})^{1/4} \nonumber \\
f_{22}(Q_1, Q_2, Q_3)&:=& \frac{5 \sqrt{6}}{48 Q_2^2}(\frac{Q_1 Q_3}{Q_2})^{1/4}- \frac{ \sqrt{6}}{16 Q_2}
(\frac{Q_3}{Q_1 Q_2})^{1/4}- \frac{ \sqrt{6}}{16 Q_2}(\frac{Q_1}{Q_3 Q_2})^{1/4} \nonumber \\
f_{33}(Q_1, Q_2, Q_3)&:=& \frac{5 \sqrt{6}}{48 Q_3^2}(\frac{Q_1 Q_2}{Q_3})^{1/4}- \frac{ \sqrt{6}}{16 Q_3}
(\frac{Q_2}{Q_1 Q_3})^{1/4}- \frac{ \sqrt{6}}{16 Q_3}(\frac{Q_1}{Q_2 Q_3})^{1/4}
\end{eqnarray}

Interestingly, it is immediate to observe that the ratio of the associated components of
statistical pair correlations vary as definite sum, which are symmetric factors of concerned
brane charges; whereas we see that there is no very direct scaling relations as in the case of
the single GH center bubbling brane configurations. Nevertheless, we notice for the distinct
$i,j,k \in \lbrace 1,2,3 \rbrace $ that the number of statistical pair correlations thus
described remains same. Moreover, we find for the multiple GH center black brane foam
configuration that the same type of  relative correlation set is followed, except that the
relative correlation functions now take realistic values over the parameters of given flux
partitions. It is worth to note that the precise scaling properties is easily visualized
just by considering the set $C_{BB}$ of the possible ratios consisting of the components of
the state-space metric tensor of the three charge bubbling black brane configurations.

Although there exist positivity of the principle components of state-space metric tensor,
but in order to accomplish local state-space stability, one needs to further demand that
all associated principle minors should be positive definite. It is rather easy to obtain
the principle minors of Hessian matrix of the entropy associated with multiple GH center
black brane foams. In fact, one finds after standard algebraic manipulations that the local
stability conditions on the one dimensional line, two dimensional surfaces and three dimensional
hyper-surfaces on the state-space manifold are respectively measured by following expressions
\begin{eqnarray}
p_1(Q_1, Q_2, Q_3)&=& -\frac{\sqrt{6}\pi}{48} Q_1^{-15/4} Q_2^{-7/4} Q_3^{-7/4}
(5 Q_1^{3/2} Q_2^{2} Q_3^{2}- 3 Q_1^{2} Q_2^{3/2} Q_3^{2}- 3 Q_1^{2} Q_2^{2} Q_3^{3/2} ) \nonumber \\
p_2(Q_1, Q_2, Q_3)&=& -\frac{\pi}{96} Q_1^{-7/2} Q_2^{-7/2} Q_3^{-3/2}
(4 Q_1^{3/2} Q_2^{2} Q_3^{2}+ Q_1^{3/2} Q_2^{2} Q_3^{3/2}- 8 Q_1^{3/2} Q_2^{3/2} Q_3^{2}  \nonumber \\ &&
- 2Q_1^{2} Q_2^{2} Q_3+ Q_1^{2} Q_2^{3/2} Q_3^{3/2}+ 4 Q_1^{2} Q_2 Q_3^{2} )
\end{eqnarray}

An investigation of definite global properties of the general bubbled black brane foam
configurations determines certain stability approximation along each directions, each planes
and each hyper-planes, as well as on the entire intrinsic state-space manifold. Specifically,
we need to determine whether the underlying three GH center foam configuration can be locally
stable on state-space planes and hyper-planes, and thus one is required to just compute the
corresponding principle minors of negative Hessian matrix of the foam entropy. Moreover, one finds
that the principle minor $p_1$ remains positive for the all $(Q_1, Q_2, Q_3)$ such that the
function $\tilde{p_1}(Q_1, Q_2, Q_3)$ satisfies
\begin{eqnarray}
\tilde{p_1}(Q_1, Q_2, Q_3):= 5 Q_1^{3/2} Q_2^{2} Q_3^{2}- 3 Q_1^{2} Q_2^{3/2} Q_3^{2}-3 Q_1^{2} Q_2^{2} Q_3^{3/2}<0
\end{eqnarray}

It is further intriguing to mention from the view-points of our present consideration that the principle
minor $p_2:= g_{11} g_{22}- g_{12}^2$ reduces to positive values for a domain of brane charges.
In particular, we see for given value of admissible brane charges that the state-space stability
on two dimensional surfaces is ensured if the function
\begin{eqnarray}
\tilde{p_2}(Q_1, Q_2, Q_3)&:=& 4 Q_1^{3/2} Q_2^{2} Q_3^{2} + Q_1^{3/2} Q_2^{2} Q_3^{3/2}
- 8 Q_1^{3/2} Q_2^{3/2} Q_3^{2} - 2Q_1^{2} Q_2^{2} Q_3 \nonumber \\ &&
+ Q_1^{2} Q_2^{3/2} Q_3^{3/2}+ 4Q_1^{2} Q_2 Q_3^{2}
\end{eqnarray}

finds definite negative value for set of given brane charges $(Q_1, Q_2, Q_3)$.
Alternatively, the linear and planer stabilities require that the given foam
configurations could be scarcely populated and thus the net brane charges are
effectively bounded by some maximum brane charges. Moreover, it is not difficult
to investigate the global stability on the full state-space configuration, which
may in fact be easily carried out by computing the determinant of the state-space
metric tensor. In this case, we observe that the determinant of the intrinsic
state-space metric tensor is a well behaved function of brane charges. From the
definition of highest principle minor, \textit{viz.}, $p_3(Q_1, Q_2, Q_3):=\Vert g \Vert$,
it is in fact not difficult to compute that the determinant of the metric tensor is
\begin{eqnarray}
\Vert g \Vert= -\frac{\pi^3 \sqrt{6}}{384} (Q_1 Q_2 Q_3)^{-13/4} \tilde{g}(Q1, Q2, Q3),
\end{eqnarray}

where the factor $ \tilde{g}(Q_1,Q_2,Q_3) $ is defined by
\begin{eqnarray}
\tilde{g}(Q_1,Q_2,Q_3)&:=& - Q_1^{3/2} Q_2 Q_3^{2} - Q_1 Q_2^{3/2}Q_3^{2}
+ 3 Q_1^{3/2} Q_2^{3/2} Q_3^{3/2}- Q_1^{3/2} Q_2^{2} Q_3  \nonumber \\ &&
- Q_1 Q_2^{2} Q_3^{3/2}- Q_1^{2} Q_2^{3/2} Q_3 + Q_1^{2} Q_2^{1/2}Q_3^{2}
+ Q_1^{2} Q_2^{2} Q_3^{1/2} \nonumber \\ && - Q_1^{2} Q_2 Q_3^{3/2}
+ Q_1^{1/2} Q_2^{2} Q_3^{2}
\end{eqnarray}

More explicitly, we see for the equal values of brane charges $Q_i:=Q$,
that the principle minors and the determinant of metric tensor being defined
as the highest principle minor, \textit{viz.}, $p_3:= g(Q)$ reduce to the
following set of values
\begin{eqnarray}
p_1(Q)= \frac{\sqrt{6}\pi}{48} Q^{-7/4}, \ 
p_2(Q)= 0, \ 
g(Q)= 0
\end{eqnarray}

Thus, the minor constraint $p_1>0$ implies that the three GH center foam configurations
under consideration are stable over the lines of the state-space manifold. However, the
vanishing of higher minor constraints, \textit{viz.}, $\{ p_i=0 \mid i= 2,3 \}$ implies
that the system is not stable over the planes and the hyper-planes of underlying state-space
configurations for any positive values of the brane charges. In particular, the constraint
$g=0$ results in an interpretation that this configuration is unstable over entire three
dimensional manifolds describing the full intrinsic state-space configuration. Moreover,
we have finally shown that the Ricci scalar curvature of the equal brane charge foam systems
indexes out of range in the division procedure, and in effect, we find that the state-space
scalar curvature function $R(Q)$ takes only vanishing arguments.

\section{Conclusion and Discussion}

We have analyzed state-space pair correlation functions and notion of stability for the
extremal and non-extremal black holes in string theory and M-theory. Our consideration
is from the viewpoints of thermodynamic state-space geometry. We find from the intrinsic
Riemannian geometry that the stability of these black branes have been divulged from
the positivity of principle minors of the space-state metric tensor. We have explicitly
analyzed the state-space configurations for (i) the two and three charge extremal black
holes, (ii) the four and six charge non-extremal black branes.

The former arises from the string theory solutions containing large number of branes
while the latter accounts for both the branes and antibranes. The numbers of
branes and antibranes offer set of parameters to define intrinsic state-space geometry.
An extension of state-space geometry is analyzed for the $D_6$-$D_4$-$D_2$-$D_0$
multi-centered black branes, small black holes with fractional electric branes, two charge
rotating fuzzy rings in the setup of Mathur's fuzzball configurations. The state-space pair
correlations and potential nature of stabilities are thereby investigated for the three charged
bubbling black brane foams. The state-space configuration finds further support from the consideration
\cite{SST, 0801.4087v1, BNTBull}, and thus the nature of state-space geometry of rotating and
non-rotating charged black branes in string theory and M-theory have respectively been examined.

In either of the black brane configurations, it has been shown that there exist an intriguing
property of relative space-state correlations that the ratio of diagonal components varies as inverse
square of the chosen parameters, while the off diagonal components vary as the inverse of the chosen
parameters. Similarly for the corresponding non-extremal configurations, we find that the ratio of
diagonal components weaken faster then the other off-diagonal components. Our analysis thus suggest
that the brane-brane statistical pair correlation functions which find asymmetric nature in comparison
with the other relative pair correlations weaken relatively faster and thus they swiftly come into an
equilibrium statistical configuration. In both the configurations, underlying microscopic notion
of the state-space interactions arise from coarse graining of the counting entropy over large number
of CFT microstates of considered black branes.

It is demonstrated that the state-space configurations arising from fluctuating spherical horizon
string theory and $M$-theory black brane solutions has as well been analyzed for finite parameter
solutions. In effect, the present paper has exemplified our out-set for diverse string theory extremal
and non-extremal black brane solutions, multi-centered black brane configurations, fuzzy rings and
single and multi Gibbon Hawking center bubbling black brane foams. It is instructive to note in this
perspective that state-space investigations of string theory and $M$-theory black brane configurations
are based on an understanding of microscopic entropy of various black branes, in which the present
consideration requires coarse graining phenomenon of large number of degenerate CFT microstates
defining an equilibrium statistical basis for chosen black brane system. The present analysis thus
offers a direct method to uncover statistical properties of fluctuating black brane configurations.

An exploration finds that the crucial ingredient in analyzing the state-space manifold of
black brane configurations depends on the parameters carried by the space-time solution or that
of an underlying microscopic conformal field theory. An illustration of state-space geometry of these
black branes includes the case of extremal and non-extremal configuration which in the either of proclaimed
configurations demonstrates that the stability constraints arising from the state-space pair correlation
functions in effect determines potential nature of the local and global correlations. It is worth to
mention that the components of a state-space metric tensor are related to the two point statistical
correlation functions which in general are intertwined with the fluctuating parameters of associated
boundary conformal field theory (CFT).

This is because that the required parameters of black brane configurations which describe the
microstates of dual conformal field theory living on the boundary may in principle be determined
via an application of AdS/CFT correspondence. Consequently, our intrinsic geometric formalism thus
described deals with an ensemble of degenerate CFT ground states which at small constant positive
temperature forms an equilibrium vacuum configuration over which we have defined the Gaussian statistical
fluctuations. It is interesting to note that the quadratic nature of Gaussian statistical fluctuations about
an equilibrium statistical configuration determines the metric tensor of associated state-space manifolds.
In either case, our explicit computation shows over definite domain of black brane parameters that the principle
components of state-space metric tensors are positive, while the non-identical off-diagonal ones may or may not.
These notions have nonetheless been explicitly observed for the case of finite electric clusters of $D_0$-$D_4$
state-space configurations that some of ratios involving off-diagonal components of metric tensor are also positive.

Interestingly, the relative correlations weakens as the concerned parameters are increased.
In particular, we find an accordance for two charge extremal black holes or an excited string with two state-space
variables, \textit{viz.}, brane numbers or brane charges and Kaluza-Klein momentum or three charge $D_1$-$D_5$-$P$
extremal solutions having $n_1$ number of $D_1$-branes, $n_5$ number of $D_5$-branes, $n_p$ number of Kaluza-Klein
momentum. Then, we find for a pair of distinct state-space variable $\{ X_i,X_j \}$ that the state-space pair
correlations of either of such extremal configurations scale as
\begin{eqnarray}
\frac{g_{ii}}{g_{jj}}= (\frac{X_j}{X_i})^2, \ 
\frac{g_{ij}}{g_{ii}}= -\frac{X_i}{X_j}
\end{eqnarray}

Furthermore, the particular behavior of generic statistical pair correlation functions characterizing
state-space configurations of four and six charge non-extremal black holes in string theory satisfy
inverse like scaling properties with integer or half-integer exponents. It may thus be envisaged that
the generic state-state correlations of string theory and $M$-theory black holes with or without rotation
decrease as an increment is made on in the parameters of concerned solution.

To appreciate definite global properties of the concerned systems, we have explained in this article
that one is required to determine nature of stabilities along each directions, each planes, and each
hyper-planes, as well as on the entire intrinsic state-space configurations. Our analysis has in effect
demonstrated that the determinant of metric tensor are negative definite as well for the configurations
having large number of branes and/ or antibrane. It has however been known from the Ruppeiner geometry that
only the classical fluctuations having definite thermal origin deal with the probability distribution
which has a positive definite invariant intrinsic Riemannian metric tensor over an equilibrium statistical
configuration. This signals that the system becomes highly quantum in nature, when all the parameters
fluctuate. In fact, our state-space construction for the string theory and M-theory black holes dealing
with the parameters of microscopic CFTs illustrates that the local stabilities, degeneracy and global
signature of a state-space manifold can as well be indefinite and in effect these notions are sensitive
to the location chosen in the moduli space geometry of the black branes.

Importantly, the sign of principle minors and determinant of the state-space metric tensor implies whether
the chosen black brane solution is thermodynamically stable or not. While, the vacuum phase transitions may
rather be characterized via the scalar curvature of concerned state-space configuration. The present
investigation thus serves as a prelude to the state-space geometry of an arbitrary parameter black brane
configuration in string theory and $M$-theory. Moreover, it has been explicates that the explored examples have
an interesting set up of intrinsic state-space geometric characterizations which are based on general nature
of the quadratic Gaussian fluctuations of chosen statistical configuration of the black branes. In this concern,
we finds in general that these configurations are categorized as

\begin{enumerate}
\item The underlying sub-configurations turn out to be well-defined over possible domains,
whenever there exist respective set of non-zero state-space principle minors.
\item The underlying full configuration turns out to be everywhere well-defined,
whenever there exist a non-zero state-space determinant.
\item The underlying configuration corresponds to an interacting statistical system,
whenever there exist a non-zero state-space scalar curvature.
\end{enumerate}

The main line of thought which has been followed up here has first been to develop an intrinsic Riemannian
geometric conception to underlying state-space geometry arising from leading order statistical interactions
which exist among various CFT microstates of (rotating) black brane configurations in string theory and $M$-theory.
The perspective notions indicates that novel scaling aspects of the state-space pair correlation and state-space
stability in effect arise from  negative Hessian matrix of the coarse graining entropy defined over an ensemble
of large number of brane microstates characterizing considered black brane attractor configurations.
Intimately, we have investigated whether the associated state-space geometries are non-degenerate and possess
non-vanishing scalar curvature imply an interacting statistical basis for these configurations, like the one
above for instance, the state-space configuration of the multi-centered $D_6$-$D_4$-$D_2$-$D_0$ solutions have
intriguingly been described, and in particular, we have presented complete list of corresponding relative
state-space correlation functions. Moreover, the stabilities of underlying configurations can as well be
perceived by noting the signs of principle minors. Similarly, the considerations of other string theory and
$M$-theory onfigurations have been divulged in this context. Interestingly, we find that the behavior of
statistical pair correlations between equilibrium microstates is governed by set of consistent parameters
defining underlying CFT vacuum configurations, and thence, the same has been anticipated to remain valid
for the other associated intrinsic geometric quantities on concerned state-space manifold, as well.

Finally, the higher order $\alpha^{\prime}$-corrections when taken into account in the underlying
effective theory are envisaged to offer diverse well-defined state-space configurations. Generically,
the $\alpha^{\prime}$-corrected state-space configurations are at least expected to be non-degenerate
than an ill-defined intrinsic Riemannian geometry arising from the leading order entropy configuration.
Such a notion has been offered for the two charge $D_0$-$D_4$ black holes or an excited string at the
leading order entropy solutions. Similar motivations along these directions have been obtained in
previous state-space investigations \cite{SST, 0801.4087v1, BNTBull}. Herewith, we have contemplated
that the state-space geometry of black branes in string theory and $M$-theory would ascribe definite
well-defined, non degenerate and curved intrinsic Riemannian manifold whose pair correlation functions
scale as inverse functions of the parameters. There are however many caveats, many things which require
further clarification and many open questions which we leave for future investigations.

\section*{Acknowledgement}

The work of S. B. has been supported in part by the European Research Council grant No. 226455,
\textit{``SUPERSYMMETRY, QUANTUM GRAVITY AND GAUGE FIELDS (SUPERFIELDS)''}.
B. N. T. would like to thank R. Gopakumar, A. Sen, R. Emparan, S. Minwalla, S. Mathur, J. Simon,
J. de Boer, B. Bhattacharjya, A. Prakash and V. Chandra for useful discussions. BNT thanks Prof.
V. Ravishankar for providing necessary support and encouragements during the preparation of the
manuscript. BNT would like to acknowledge the hospitality of INFN-LNF, Frascati, Italy where part
of this work was performed.

\section*{Appendix}
In this appendix, we provide explicit forms of the state-space relative correlation
functions at the first and second centers of the multicentered $D_6D_4D_2D_0$ black holes
describing family of the four charged configurations. Our analysis illustrates that the
physical properties of the specific state-space correlations may exactly be exploited in
general. The definite behavior of state-space correlations as accounted in the concerned section
suggests that the various intriguing single center and multi-center state-space examples of
black brane solutions includes nice properties that they do have definite stability properties,
except that the determinant may be non-positive definite in some cases. As mentioned in the
main sections, these $D_6D_4D_2D_0$ configurations are generically well-defined and indicate
an interacting statistical basis. We discover here that their state-space geometries indicate
possible nature of general two center equilibrium thermodynamic configurations. Significantly,
we notice from the very definition of intrinsic metric tensor that the relative state-space
correlations may be analyzed as follows.

\subsection*{Appendix (A): State-space relative correlations at the first center}

Here, we shall explicitly provide the exact expressions for the four parameter
multi-centered solutions at the first center of the double centered black holes.
It turns out that the functional nature of large number of branes within a small
neighborhood of statistical fluctuations introduced in an equilibrium ensemble
of brane configurations may precisely be divulged. Surprisingly, we can expose
in this framework that the relative state-space correlations at the first center
with charges $p0:=1$; $p:=3\Lambda$; $q:=6\Lambda^2$; and $q0:=-6\Lambda$ take
an exact and simple set of expressions

\begin{eqnarray}
c_{1112}&=& -2 \frac{ 6 \Lambda^2+ 12 \Lambda^4+ 8 \Lambda^6+ 1}
{7 \Lambda^2+ 16 \Lambda^4+ 12 \Lambda^6+ 1}, \ 
c_{1113}= 2 \frac{6 \Lambda^2+ 12 \Lambda^4+ 8 \Lambda^6+ 1}
{4 \Lambda^2+ 4 \Lambda^4+ 1} \nonumber\\
c_{1114}&=& -108 \frac{6 \Lambda^2+ 12 \Lambda^4+ 8 \Lambda^6+ 1}
{18 \Lambda^4+ 27 \Lambda^6- 1}, \ 
c_{1122}= 6 \Lambda^2 \frac{6 \Lambda^2+ 12 \Lambda^4+ 8 \Lambda^6+ 1}
{13 \Lambda^2+ 30 \Lambda^4+ 2+ 24 \Lambda^6}\nonumber\\
c_{1123}&=& -36 \Lambda^3 \frac{6 \Lambda^2+ 12 \Lambda^4+ 8 \Lambda^6 + 1}
{42 \Lambda^4+ 12 \Lambda^2+ 45 \Lambda^6+ 1}, \ 
c_{1124}= 12  \frac{6 \Lambda^2+ 12 \Lambda^4+ 8 \Lambda^6 + 1}
{1+ 2 \Lambda^2} \nonumber\\
c_{1133}&=& 36 \Lambda^2 \frac{6 \Lambda^2+ 12 \Lambda^4+ 8 \Lambda^6 + 1}
{2+ 9\Lambda^2+ 12\Lambda^4}, \ 
c_{1134}= -72 \Lambda \frac{6 \Lambda^2+ 12 \Lambda^4+ 8 \Lambda^6 + 1}
{1+ 3 \Lambda^2} \nonumber\\
c_{1144}&=& 1296 \Lambda^2 + 2592 \Lambda^4 + 1728 \Lambda^6 + 216, \ 
c_{1213}= - \frac{7 \Lambda^2 + 16 \Lambda^4 + 1 + 12 \Lambda^6}
{\Lambda(4\Lambda^2 + 1 + 4\Lambda^4)} \nonumber\\
c_{1222}&=& - 3\Lambda \frac{7 \Lambda^2 + 16 \Lambda^4 + 1 + 12 \Lambda^6}
{13 \Lambda^2+ 30\Lambda^4+ 2+ 24 \Lambda^6}, \ 
c_{1223}= 18 \Lambda^2 \frac{7 \Lambda^2 + 16 \Lambda^4 + 1 + 12 \Lambda^6}
{42 \Lambda^4+ 12 \Lambda^2+ 45 \Lambda^6+ 1} \nonumber\\
c_{1224}&=& -6 \frac{7 \Lambda^2 + 16 \Lambda^4 + 1 + 12 \Lambda^6}
{\Lambda(1 + 2\Lambda^2)}, \ 
c_{1233}= -18\Lambda \frac{7 \Lambda^2 + 16 \Lambda^4 + 1 + 12 \Lambda^6}
{9 \Lambda^2 + 12 \Lambda^4 + 2} \nonumber\\
c_{1234}&=& 36 \frac{7 \Lambda^2 + 16 \Lambda^4 + 1 + 12 \Lambda^6}
{3 \Lambda^2 + 1}, \ 
c_{1244}= -108 \frac{7 \Lambda^2 + 16 \Lambda^4 + 1 + 12 \Lambda^6}
{\Lambda} \nonumber\\
c_{1314}&=& -54 \Lambda^3 \frac{ 4 \Lambda^2 + 1 + 4 \Lambda^4}
{18 \Lambda^4 + 27 \Lambda^6- 1}, \ 
c_{1322}= 3 \Lambda^2 \frac{4 \Lambda^2 + 1 + 4 \Lambda^4}
{13 \Lambda^2 + 30 \Lambda^4 + 2 + 24 \Lambda^6} \nonumber\\
c_{1323}&=& -18 \Lambda^3 \frac{4 \Lambda^2 + 1 + 4 \Lambda^4 }
{42 \Lambda^4 + 12 \Lambda^2 + 45 \Lambda^6 + 1}, \ 
c_{1324}= 6 \frac{4 \Lambda^2 + 1 + 4 \Lambda^4 }
{1+ 2\Lambda^2} \nonumber\\
c_{1333}&=& 18 \Lambda^2 \frac{4 \Lambda^2 + 1 + 4 \Lambda^4}
{9 \Lambda^2 + 12 \Lambda^4 + 2}, \ 
c_{1334}= -36 \Lambda \frac{4 \Lambda^2 + 1 + 4 \Lambda^4}
{1+ 3\Lambda^2} \nonumber\\
c_{1344}&=& 432 \Lambda^2 + 108 + 432 \Lambda^4, \ 
c_{1422}= -\frac{1}{18 \Lambda} \frac{18 \Lambda^4 + 27 \Lambda^6- 1}
{13 \Lambda^2 + 30 \Lambda^4 + 2 + 24 \Lambda^6} \nonumber\\
c_{1423}&=& \frac{1}{3} \frac{18 \Lambda^4 + 27 \Lambda^6- 1}
{42 \Lambda^4 + 12 \Lambda^2 + 45 \Lambda^6 + 1}, \ 
c_{1424}= -\frac{1}{9 \Lambda^3} \frac{18 \Lambda^4 + 27 \Lambda^6- 1}
{1 + 2 \Lambda^2} \nonumber\\
c_{1433}&=& -\frac{1}{3 \Lambda} \frac{18 \Lambda^4 + 27 \Lambda^6- 1}
{9 \Lambda^2 + 12 \Lambda^4 + 2}, \ 
c_{1434}= \frac{2}{3 \Lambda^2} \frac{18 \Lambda^4 + 27 \Lambda^6- 1}
{1+ 3\Lambda^2} \nonumber\\
c_{1444}&=& -\frac{2}{\Lambda^3} (18 \Lambda^4 + 27 \Lambda^6- 1), \ 
c_{2223}= -6 \Lambda \frac{13 \Lambda^2+ 30 \Lambda^4+ 24 \Lambda^6+ 2}
{42 \Lambda^4+ 12 \Lambda^2+ 45 \Lambda^6+ 1} \nonumber\\
c_{2224}&=&  \frac{2}{\Lambda^2} \frac{13 \Lambda^2+ 30 \Lambda^4+ 24 \Lambda^6+ 2}
{1 + 2 \Lambda^2}, \ 
c_{2233}= 6 \frac{13 \Lambda^2+ 30 \Lambda^4+ 24 \Lambda^6+ 2}
{9 \Lambda^2 + 12 \Lambda^4 + 2} \nonumber\\
c_{2234}&=& -12 \frac{13 \Lambda^2+ 30 \Lambda^4+ 24 \Lambda^6+ 2}
{\Lambda(1+ 3\Lambda^2)}, \ 
c_{2244}= 36 \frac{13 \Lambda^2+ 30 \Lambda^4+ 24 \Lambda^6+ 2}
{\Lambda^2} \nonumber\\
c_{2324}&=& -\frac{1}{3 \Lambda^3} \frac{42 \Lambda^4 + 12 \Lambda^2 + 45 \Lambda^6 + 1}
{1 + 2 \Lambda^2}, \ 
c_{2333}= -\frac{1}{\Lambda} \frac{42 \Lambda^4 + 12 \Lambda^2 + 45 \Lambda^6 + 1}
{9 \Lambda^2 + 12 \Lambda^4 + 2} \nonumber\\
c_{2334}&=& \frac{2}{\Lambda^2} \frac{42 \Lambda^4 + 12 \Lambda^2 + 45 \Lambda^6 + 1}
{1 + 3 \Lambda^2}, \ 
c_{2344}= -\frac{6}{\Lambda^2} (42 \Lambda^4 + 12 \Lambda^2 + 45 \Lambda^6 + 1) \nonumber\\
c_{2433}&=& 3\Lambda^2 \frac{1 + 2 \Lambda^2}
{9 \Lambda^2 + 12 \Lambda^4 + 2}, \ 
c_{2434}= -6 \Lambda \frac{1 + 2 \Lambda^2}
{1 + 3 \Lambda^2} \nonumber\\
c_{2444}&=& 18 + 36 \Lambda^2, \ 
c_{3334}= -\frac{2}{\Lambda} \frac{9 \Lambda^2 + 12 \Lambda^4 + 2}
{3 \Lambda^2 + 1}  \nonumber\\
c_{3344}&=& \frac{6}{\Lambda^2} (9 \Lambda^2 + 12 \Lambda^4 + 2), \ 
c_{3444}= -\frac{3}{\Lambda} (3 \Lambda^2 + 1)
\end{eqnarray}

\subsection*{Appendix (B): State-space relative correlations at the second center}

As stated earlier, the state-space metric in the multi-centered and single black brane
configurations is given by negative Hessian matrix of concerned entropy. Here,
the charges on the branes in a given configuration are respected to be extensive
variables. In this case, we find that the four distinct large charges characterize
the intrinsic state-space correlation functions. In fact, our computation shows
that the exact set of correlations at the second center are given by employing
the previously defined notations. We have similarly presented for the second center
of the $D_6$-$D_4$-$D_2$-$D_0$ system that the relative state-space correlations
simplify

\begin{eqnarray}
c_{1112}&=& -2 \frac{6 \Lambda^2+ 12 \Lambda^4+ 8 \Lambda^6+ 1}
{7 \Lambda^2+ 16 \Lambda^4+ 12 \Lambda^6+ 1}, \ 
c_{1113}= 2 \frac{6 \Lambda^2+ 12 \Lambda^4+ 8 \Lambda^6+ 1}
{4 \Lambda^2+ 4 \Lambda^4+ 1} \nonumber\\
c_{1114}&=& 108\Lambda^3 \frac{6 \Lambda^2+ 12 \Lambda^4+ 8 \Lambda^6+ 1}
{18 \Lambda^4+ 27 \Lambda^6- 1}, \ 
c_{1122}= 6 \Lambda^2 \frac{6 \Lambda^2+ 12 \Lambda^4+ 8 \Lambda^6+ 1}
{13 \Lambda^2+ 30 \Lambda^4+ 2+ 24 \Lambda^6} \nonumber\\
c_{1123}&=& 36 \Lambda^3 \frac{6 \Lambda^2+ 12 \Lambda^4+ 8 \Lambda^6+ 1}
{42 \Lambda^4+ 12 \Lambda^2+ 45 \Lambda^6+ 1}, \ 
c_{1124}= 12  \frac{6 \Lambda^2+ 12 \Lambda^4+ 8 \Lambda^6 + 1}
{1+ 2 \Lambda^2} \nonumber\\
c_{1133}&=& 36 \Lambda^2 \frac{6 \Lambda^2+ 12 \Lambda^4+ 8 \Lambda^6 + 1}
{2+ 9\Lambda^2+ 12\Lambda^4}, \ 
c_{1134}= -72 \Lambda \frac{6 \Lambda^2+ 12 \Lambda^4+ 8 \Lambda^6 + 1}
{1+ 3 \Lambda^2} \nonumber\\
c_{1144}&=& 1296 \Lambda^2 + 2592 \Lambda^4 + 1728 \Lambda^6 + 216, \ 
c_{1213}= - \frac{7 \Lambda^2 + 16 \Lambda^4 + 1 + 12 \Lambda^6}
{\Lambda(4\Lambda^2 + 1 + 4\Lambda^4)} \nonumber\\
c_{1222}&=& -3 \Lambda \frac{7 \Lambda^2 + 16 \Lambda^4 + 1 + 12 \Lambda^6}
{13 \Lambda^4 + 12 \Lambda^2+ 45\Lambda^6+ 1}, \ 
c_{1223}= -18 \Lambda^2 \frac{7 \Lambda^2 + 16 \Lambda^4 + 1 + 12 \Lambda^6}
{42 \Lambda^2 + 12\Lambda^2+ 45\Lambda^6+ 1}  \nonumber\\
c_{1224}&=& -6 \frac{7 \Lambda^2 + 16 \Lambda^4 + 1 + 12 \Lambda^6}
{\Lambda(1 + 2\Lambda^2)}, \ 
c_{1233}= -18\Lambda \frac{7 \Lambda^2 + 16 \Lambda^4 + 1 + 12 \Lambda^6}
{9 \Lambda^2 + 12 \Lambda^4 + 2}  \nonumber\\
c_{1234}&=& 36 \frac{7 \Lambda^2 + 16 \Lambda^4 + 1 + 12 \Lambda^6}
{3 \Lambda^2 + 1}, \ 
c_{1244}= -108 \frac{7 \Lambda^2 + 16 \Lambda^4 + 1 + 12 \Lambda^6}
{\Lambda}  \nonumber\\
c_{1314}&=& 54 \Lambda^3 \frac{ 4 \Lambda^2 + 1 + 4 \Lambda^4}
{18 \Lambda^4 + 27 \Lambda^6- 1}, \ 
c_{1322}= 3 \Lambda^2 \frac{4 \Lambda^2 + 1 + 4 \Lambda^4}
{13 \Lambda^2 + 30 \Lambda^4 + 2 + 24 \Lambda^6}  \nonumber\\
c_{1323}&=& 18 \Lambda^3 \frac{4 \Lambda^2 + 1 + 4 \Lambda^4 }
{42 \Lambda^4 + 12 \Lambda^2 + 45 \Lambda^6 + 1}, \ 
c_{1324}= 6 \frac{4 \Lambda^2 + 1 + 4 \Lambda^4 }
{1+ 2\Lambda^2}   \nonumber\\
c_{1333}&=& 18 \Lambda^2 \frac{4 \Lambda^2 + 1 + 4 \Lambda^4}
{9 \Lambda^2 + 12 \Lambda^4 + 2}, \ 
c_{1334}= -36 \Lambda \frac{4 \Lambda^2 + 1 + 4 \Lambda^4}
{1+ 3\Lambda^2} \nonumber\\
c_{1344}&=& 432 \Lambda^2 + 108 + 432 \Lambda^4, \ 
c_{1422}= \frac{1}{18 \Lambda} \frac{18 \Lambda^4 + 27 \Lambda^6- 1}
{13 \Lambda^2 + 30 \Lambda^4 + 2 + 24 \Lambda^6} \nonumber\\
c_{1423}&=& \frac{1}{3} \frac{18 \Lambda^4 + 27 \Lambda^6- 1}
{42 \Lambda^4 + 12 \Lambda^2 + 45 \Lambda^6 + 1}, \ 
c_{1424}= \frac{1}{9 \Lambda^3} \frac{18 \Lambda^4 + 27 \Lambda^6- 1}
{1 + 2 \Lambda^2}  \nonumber\\
c_{1433}&=& \frac{1}{3 \Lambda} \frac{18 \Lambda^4 + 27 \Lambda^6- 1}
{9 \Lambda^2 + 12 \Lambda^4 + 2}, \ 
c_{1434}= -\frac{2}{3 \Lambda^2} \frac{18 \Lambda^4 + 27 \Lambda^6- 1}
{1+ 3\Lambda^2} \nonumber\\
c_{1444}&=& \frac{2}{\Lambda^3} (18 \Lambda^4 + 27 \Lambda^6- 1), \ 
c_{2223}=  6\Lambda \frac{13 \Lambda^2+ 30 \Lambda^4+ 24 \Lambda^6+ 2}
{42 \Lambda^4+ 12 \Lambda^2+ 45 \Lambda^6+ 1}  \nonumber\\
c_{2224}&=& \frac{2}{\Lambda^2} \frac{13 \Lambda^2+ 30 \Lambda^4+ 24 \Lambda^6+ 2}
{1 + 2 \Lambda^2}, \ 
c_{2233}= 6 \frac{13 \Lambda^2+ 30 \Lambda^4+ 24 \Lambda^6+ 2}
{9 \Lambda^2 + 12 \Lambda^4 + 2}  \nonumber\\
c_{2234}&=& -12 \frac{13 \Lambda^2+ 30 \Lambda^4+ 24 \Lambda^6+ 2}
{\Lambda(1+ 3\Lambda^2)}, \ 
c_{2244}= 36 \frac{13 \Lambda^2+ 30 \Lambda^4+ 24 \Lambda^6+ 2}
{\Lambda^2}  \nonumber\\
c_{2324}&=& \frac{1}{3 \Lambda^3} \frac{42 \Lambda^4 + 12 \Lambda^2 + 45 \Lambda^6 + 1}
{1 + 2 \Lambda^2}, \ 
c_{2333}= \frac{1}{\Lambda} \frac{42 \Lambda^4 + 12 \Lambda^2 + 45 \Lambda^6 + 1}
{9 \Lambda^2 + 12 \Lambda^4 + 2}  \nonumber\\
c_{2334}&=& -\frac{2}{\Lambda^2} \frac{42 \Lambda^4 + 12 \Lambda^2 + 45 \Lambda^6 + 1}
{1 + 3 \Lambda^2}, \ 
c_{2344}= \frac{6}{\Lambda^2} (42 \Lambda^4 + 12 \Lambda^2 + 45 \Lambda^6 + 1)  \nonumber\\
c_{2433}&=& 3\Lambda^2 \frac{1 + 2 \Lambda^2}
{9 \Lambda^2 + 12 \Lambda^4 + 2}, \ 
c_{2434}= -6 \Lambda \frac{1 + 2 \Lambda^2}
{1 + 3 \Lambda^2}  \nonumber\\
c_{2444}&=& 18 + 36 \Lambda^2, \ 
c_{3334}= -\frac{2}{\Lambda} \frac{9 \Lambda^2 + 12 \Lambda^4 + 2}
{3 \Lambda^2 + 1}  \nonumber\\
c_{3344} &=&\frac{6}{\Lambda^2} (9 \Lambda^2 + 12 \Lambda^4 + 2), \ 
c_{3444}= -\frac{3}{\Lambda} (3 \Lambda^2 + 1)
\end{eqnarray}

\end{document}